\documentclass[lettersize,journal]{IEEEtran}
\usepackage{amsthm}
\usepackage{amsmath,amsfonts}
\usepackage{algorithmic}
\usepackage{algorithm}
\usepackage{amssymb}
\usepackage{array}
\usepackage[caption=false,font=normalsize,labelfont=sf,textfont=sf]{subfig}
\usepackage{textcomp}
\usepackage{stfloats}
\usepackage{url}
\usepackage{verbatim}
\usepackage{graphicx}
\usepackage{cite}
\usepackage{cases}
\usepackage{hyperref}
\usepackage{bm}
\usepackage{color}

\hypersetup{colorlinks=true,
	linkcolor=blue,
	citecolor=blue,      
	urlcolor=black,
}

\newtheoremstyle{mylemma}
{}{}                 
{\normalfont}        
{}                   
{\bfseries}         
{.}                  
{ }                  
{\thmname{#1}\thmnumber{ #2}\thmnote{ (#3)}} 

\theoremstyle{mylemma}
\newtheorem{lemma}{Lemma}
\newtheorem{remark}{Remark}
\newtheorem{theorem}{Theorem}

\newtheorem{proposition}{Proposition}

\begin{document}

\title{Spectral and Energy Efficiency Tradeoff for Pinching-Antenna Systems}

\author{Zihao Zhou,~\IEEEmembership{Graduate Student Member,~IEEE}, Zhaolin Wang,~\IEEEmembership{Member,~IEEE}, and Yuanwei Liu,~\IEEEmembership{Fellow,~IEEE}
\thanks{The authors are with the Department of Electrical and Computer Engineering, The University of Hong Kong, Hong Kong (e-mail: eezihaozhou@connect.hku.hk,zhaolin.wang@hku.hk,yuanwei@hku.hk)}}

\maketitle

\begin{abstract}
The joint transmit and pinching beamforming design for spectral efficiency (SE) and energy efficiency (EE) tradeoff in pinching-antenna systems (PASS) is proposed, under practical channel and energy consumption models. In the single-user scenario, it is proved that the optimal pinching antenna (PA) positions are independent of the transmit beamforming. Based on this insight, a two-stage joint beamforming design is proposed. Specifically, in the first stage, a general PA placement framework is proposed for multi-waveguide systems. In the second stage, the closed-form solution for the optimal transmit beamformer is derived given the optimized PA positions. In the multi-user scenario, an alternating optimization (AO)-based joint beamforming design is proposed to balance the SE-EE performance while taking the quality-of-service (QoS) requirements into account. It is proved that the proposed AO-based algorithm is guaranteed to converge when no constraints are violated in PA placement subproblem. Numerical results demonstrate that: 1) the proposed algorithms effectively improve joint SE-EE performance; 2) PASS exhibits strong robustness against variations in the service area along the waveguide direction.
\end{abstract}

\section{Introduction}
\IEEEPARstart{T}{he} pursuit of higher capacity and efficiency has long been a consistent goal in wireless communication systems\cite{liu2024survey}. Over the past few decades, guided by the Shannon formula, researchers have optimized the systems by focusing on bandwidth\cite{paulraj2004overview}, transmit power\cite{islam2017power}, and even the noise\cite{basar2023noise}, while treating the wireless channel as a fixed system parameter. In recent years, however, the emergence of flexible-antenna techniques, such as reconfigurable intelligent surfaces (RISs)\cite{liu2021reconfigurable}, fluid-antennas\cite{wong2021fluid} and movable-antennas\cite{zhu2024modeling} has unlocked new possibilities for the design of wireless communication systems by treating the wireless channel itself as a reconfigurable parameter. Specifically, RISs modify the channel via programmable phase shifters, whereas fluid- and movable-antenna systems physically adjust the position of antenna elements to establish favorable channel conditions\cite{shan2025exploiting}. More importantly, flexible-antenna techniques are inherently compatible with multiple-input multiple-output (MIMO), a technology that has been among the most celebrated ones over the past thirty years. In particular, they introduce a new layer of electromagnetic beamforming through antenna reconfigurability within the MIMO architecture\cite{heath2025tri}. This intrinsic compatibility undoubtedly injects strong vitality into the development of flexible-antenna techniques.

Despite recent advances, the existing flexible-antenna techniques face inherent limitations, which may hinder their further application. Firstly, they exhibit insufficient capacities against large-scale fading. For instance, RIS introduces two separate links: the transmitter-to-RIS link and the RIS-to-receiver link. The increased propagation distance leads to more severe path loss\cite{ding2025pinching}. While for fluid/movable antennas, the movement of antenna elements is constrained to the wavelength scale\cite{xu2025rate}, which typically has an insignificant impact on large-scale fading. Secondly, for the flexible-antenna systems mentioned above, once deployed, it is difficult to change the number of antenna elements anymore, which means that the array size cannot be adaptively modified according to different user demands. This reflects the ``inflexibility'' of the existing flexible-antenna systems.

To address the aforementioned challenges, pinching-antenna system (PASS) emerges as a novel flexible-antenna technique to overcome large-scale fading while offering ``ultra-flexibility''. The concept of PASS and the experimental demonstrations were first given by NTT DOCOMO\cite{fukuda2022pinchng}. Specifically, PASS consists of one or more dielectric waveguides as transmission medium. By deploying multiple small, separated dielectric elements, referred to as \emph{pinching antenna (PA)}, onto the waveguides, the in-waveguide radio waves can be radiated into free space via these PAs. Unlike traditional flexible-antenna systems, PASS distinguishes itself by ``ultra-flexibility'', which mainly includes two aspects: 1) PASS overcomes the restriction that antenna movement can only be limited to the wavelength scale. Specifically, with the aid of waveguides which span across several to tens of meters\cite{shan2025exploiting}, PAs can be deployed close to users, substantially reducing large-scale path loss and alleviating blockage from obstacles. Moreover, by jointly optimizing the placement of multiple PAs, a new capability, which is referred to as \emph{pinching beamforming}, is enabled to further enhance signal quality; 2) by simply attaching additional PAs or releasing existing ones, changing the array size in PASS becomes extremely straightforward, which can well support the on-demand deployment.

These advantages aforementioned have spurred growing interest in investigating PASS from the perspectives of theoretical analysis and system optimization. Specifically, the authors of \cite{ding2025pinching} presented the first comprehensive theoretical analysis for PASS in terms of ergodic sum rate. In \cite{tyrovolas2025performance}, the closed-form expressions for the outage probability and average rate of PASS were derived. The authors of \cite{ouyang2025array} investigated PASS from the perspective of array gain where the closed-form upper bound on the array gain is derived, unveiling the optimal number of PAs and spacing. Considering PASS-assisted uplink communication system, the closed-form expressions for analytical, asymptotic and approximated ergodic rate were derived in \cite{hou2025on}. In \cite{ding2025analytical}, the optimal position of PA was given under the user-fairness-oriented orthogonal multiple access (OMA) based communication. Furthermore, a growing body of recent literature is devoted to the performance analysis of PASS in combination with various other emerging technologies, such as simultaneous wireless information and power transfer (SWIPT)\cite{zhang2025on}, integrated sensing and communications (ISAC)\cite{ouyang2025rate} and non-orthogonal multiple access (NOMA)\cite{cheng2025on}.

In addition to performance analysis, researchers have also focused on optimizing the performance of PASS-assisted wireless communication systems in more general settings. For the single-waveguide employment, the authors of \cite{xu2025rate}, proposed a two-stage algorithm to optimize the positions of PAs with the aim at maximizing the rate of the user. The sum rate maximization problem was investigated in PASS-assisted NOMA systems in \cite{zhou2025sum,zeng2025sum,hu2025sum}. Consider discrete PAs' positions, the authors of \cite{wang2025antenna} investigated how many and at which positions the PAs need to be activated such that the sum rate is maximized. Also focusing on sum rate maximization, the authors of \cite{zhang2025uplink} addressed this problem in PASS-assisted uplink multiuser multiple-input single-output (MISO) systems. In \cite{li2025sum}, the sum rate was maximized by jointly optimizing the resource allocation and PAs' positions in a PASS-assisted wireless powered communication network (WPCN). In \cite{zhao2025waveguide}, a novel concept of waveguide division multiple access (WDMA) was proposed, and the power allocation as well as the pinching beamforming were jointly designed to maximized the sum rate.

Apart from the aforementioned works focusing on spectral efficiency (SE) optimization\cite{xu2025rate,zhou2025sum,zeng2025sum,hu2025sum,wang2025antenna,zhang2025uplink,li2025sum,zhao2025waveguide}, there have been other studies investigating the energy efficiency (EE) of PASS, which is an important metric that takes into account both system power consumption and throughput. For example, in \cite{zeng2025energy}, a NOMA-assisted uplink single-waveguide PASS was investigated, with the objective to maximize the EE by jointly optimizing the users' transmit power and the positions of PAs. The authors of \cite{zeng2025energy2} shifted the attention to downlink scenarios in a PASS assisted time-division multiple access (TDMA)-based system by jointly optimizing the transmit power, time allocation, as well as the positions of PAs under the quality-of-service (QoS) constraint.

While the work introduced above demonstrates the effectiveness of PASS in enhancing both SE and EE, they have predominantly focused on the optimization of either SE or EE. However, the objectives of maximizing SE and EE do not always coincide. On the contrary, they sometimes conflict, especially in moderate and high signal-to-noise ratio (SNR) regimes\cite{zhou2022rate}. It is therefore essential to analyze and optimize the SE-EE trade-off for PASS, which is a critical aspect that remains largely unexplored. The main contributions are summarized as follows:
\begin{itemize}
	\item We investigate the SE-EE trade-off for PASS, both in single- and multi-user scenarios under practical channel and energy consumption models. A transmit and pinching beamforming optimization problem is formulated to maximize the joint SE-EE performance while satisfying the QoS requirements, transmit power budget, and positions constraints of PAs. In contrast to conventional SE or EE-targeted design for PASS, our approach offers a generic framework to strike a balance between SE and EE, which provides guidance for configuring PASS flexibly.
	\item For single-user scenario, we prove that the optimal PA positions are independent of the transmit beamforming but not vice versa. Based on this insight, an efficient two-stage joint beamforming design is proposed: in the first stage, we propose a general PA placement framework that operates through a coarse-to-fine adjustment process; In the second stage, under maximum ratio transmission (MRT), a closed-form solution is derived for the optimal transmit power with the given optimized PA positions.
	\item For multi-user scenario, an alternating optimization (AO) framework is proposed to address the joint SE-EE maximization problem. Using zero-forcing (ZF) beamforming, the optimization of transmit beamformer reduces to the design of a diagonal scalar power control matrix, which is optimized by deriving the convex upper bounds of the constraints. Meanwhile, the pinching beamforming is optimized by element-wise sequential partial swarm optimization (PSO) method. Furthermore, we prove that the proposed AO-based algorithm is guaranteed to converge when no constraints are violated in PA placement subproblem.
	\item Extensive numerical results validate the effectiveness of the proposed algorithms. The results demonstrate that: 1) the proposed algorithm effectively improve the joint SE-EE performance with a small number of iterations; 2) PASS exhibits strong robustness against variations in the service area along the waveguide direction due to the flexible placement of PAs. These results confirm the scalability and potential of PASS for communication systems.
\end{itemize}
\vspace{-0.15em}
The rest of this paper is structured as follows. In Section \ref{sec:2}, a joint SE-EE design in PASS-enabled single-user scenario is investigated. In Section \ref{sec:3}, the transmit and pinching beamforming are optimized in the multi-user scenario. Numerical results are provided in Section \ref{sec:numerical results}, which is followed by our conclusions in Section \ref{sec:conclusion}.

\emph{Notations:} Scalars, vectors, and matrices are represented by regular, bold lowercase, and bold uppercase (e.g., x, $\bm{{\rm x}}$ and $\bm{{\rm X}}$) letters, respectively. The set of complex and real numbers are denoted by $\mathbb{C}$ and $\mathbb{R}$, respectively. The inverse, transpose, conjugate transpose, and trace operators are denoted by $(\cdot)^{-1}$, $(\cdot)^T$, $(\cdot)^H$, and ${\rm tr}(\cdot)$, respectively. The absolute value and Euclidean norm are denoted by $\vert\cdot\vert$ and $\Vert\cdot\Vert$, respectively. $\mathcal{CN}(a,b^2)$ is denoted as a circularly symmetric complex Gaussian distribution with mean $a$ and variance $b^2$. The expectation operator is denoted by $\mathbb{E}[\cdot]$. 

\section{Joint SE-EE Design For Single-User PASS}\label{sec:2}
\subsection{System Model}\label{system_model_in_single_user_PASS}
\begin{figure}[t]
	\centering
	\includegraphics[width=3.5in]{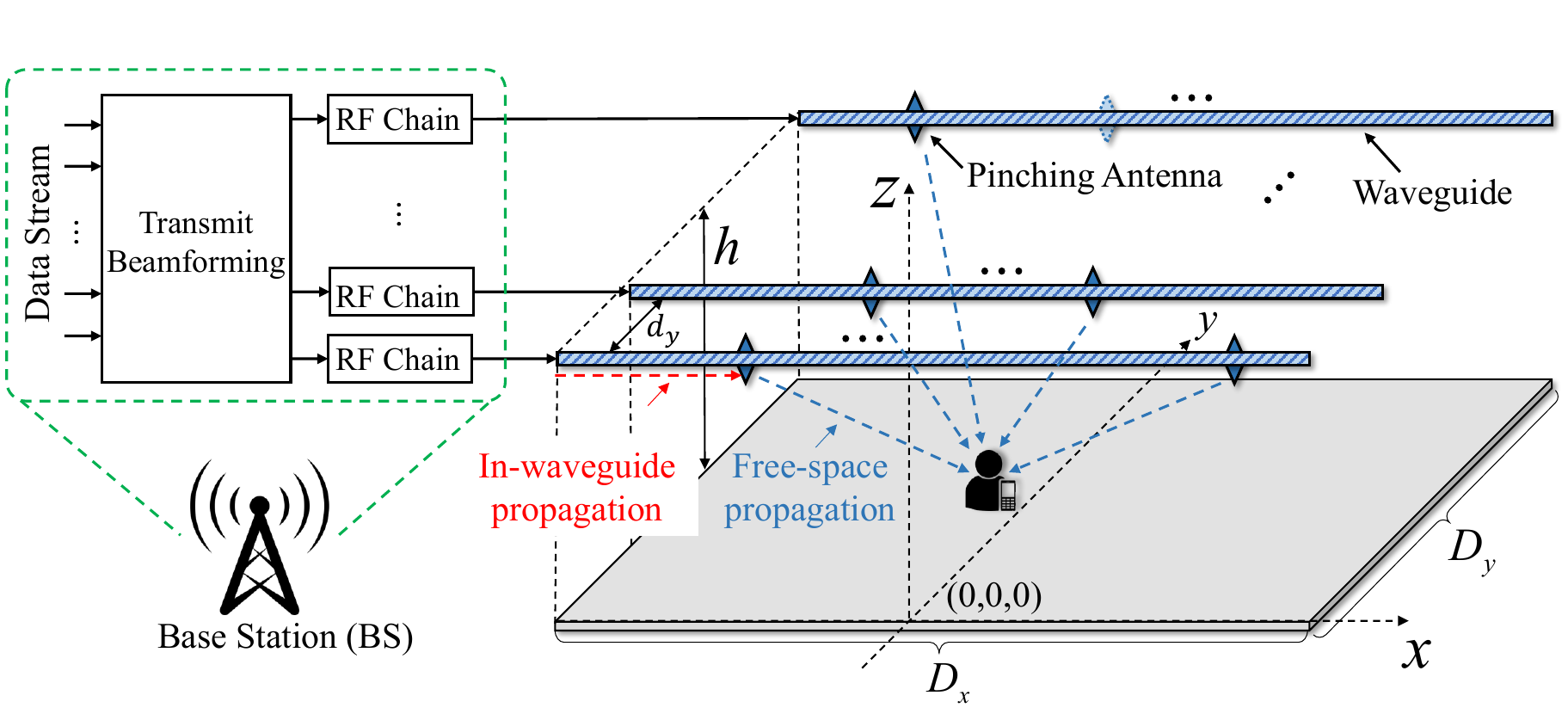}
	\caption{System model for PASS-enabled single-user communications.}
	\label{figure_1}
\end{figure}

We first focus on a single-user downlink PASS as illustrated in Fig. \ref{figure_1}. A single-antenna user is randomly located in a region of size $D=D_{{\rm x}} \times D_{{\rm y}}$ m$^2$, with its position being specified as $\bm{\psi}^u=[x^u, y^u, 0]^{\rm T}$. A base station (BS) is equipped with $M$ dielectric waveguides, each is attached with $N$ pinching antennas (PAs), to simultaneously support this user. We assume that the waveguides are deployed along the $x$-axis at an altitude of $h$, and arranged in an array along the $y$-axis, with inter-waveguide spacing $d_y=D_y/(M-1)$ m. Let $\mathcal{M}$ denotes the set of waveguides, and $\mathcal{N}_m$ the set of PAs mounted on the $m$-th waveguide. Each waveguide is fed by a dedicated radio frequency (RF) chain located at $\bm{\alpha}_{m,0}=[-D_x/2, y_m, h]^{\rm T}$ for $m\in \mathcal{M}$. The position of the $n$-th PA on the $m$-th waveguide is given by $\bm{\alpha}_{m,n}=[x_{m,n}^p, y_m, h]^{\rm T}$ for $m\in \mathcal{M}$ and $n\in \mathcal{N}_m$. The $x$-axis coordinates of all PAs can be stacked into a matrix $\bm{{\rm X}}=[\bm{{\rm x}}_1, \bm{{\rm x}}_2, ..., \bm{{\rm x}}_M]\in \mathbb{R}^{N\times M}$, where $\bm{{\rm x}}_m=[x_{m,1}^p,x_{m,2}^p,...,x_{m,N}^p]^{{\rm T}}\in \mathbb{R}^{N\times 1}$ represents the $x$-axis coordinate vector of PAs on the $m$-th waveguide. In this work, $\bm{{\rm X}}$ is treated as a key optimization variable for the design of pinching beamforming.

In PASS, the end-to-end channel from the BS to ground user consists of two components: \textit{in-waveguide propagation} and \textit{free-space propagation}\cite{liu2025pinching}, as shown in Fig. \ref{figure_1}. The former consists of in-waveguide attenuation\cite{khalili2025pinching, papanikolaou2025resolving} and phase shift. The in-waveguide channel vector for the $m$-th waveguide can be expressed as
\begin{equation}\label{sec2A_eq1}
	\bm{{\rm g}}_m(\bm{{\rm x}}_m)\!\!=\!\!\left[\omega^m_{1}\!{\rm e}^{-j \frac{2\pi}{\lambda_g}\Vert \bm{\alpha}_{m,0}\!-\bm{\alpha}_{m,1}\Vert},\!\cdots\!,\omega^m_{N}{\rm e}^{-j \frac{2\pi}{\lambda_g}\Vert \bm{\alpha}_{m,0}-\!\bm{\alpha}_{m,N}\Vert}\right]^{\rm T}\!\!\!\!,
\end{equation}
where $\omega^m_n={\rm e}^{-a\Vert \bm{\alpha}_{m,0}-\bm{\alpha}_{m,n}\Vert}, n=1,2,\cdots,N$ represents the in-waveguide attenuation with $a$ being the attenuation coefficient, $\lambda_g=\lambda/n_{{\rm eff}}$ denotes the guided wavelength with $\lambda$ and $n_{{\rm eff}}$ being the free-space wavelength and the waveguide effective refractive index, respectively. Therefore, the overall in-waveguide channel matrix can be defined by block-diagonal matrix $\bm{{\rm G}}(\bm{{\rm X}})\in \mathbb{C}^{MN\times M}$, which is given as
\begin{equation}\label{sec2A_eq2}
	\bm{{\rm G}}(\bm{{\rm X}})= {\rm blkdiag}
	\left\{\bm{{\rm g}}_1(\bm{{\rm x}}_1),...,\bm{{\rm g}}_M(\bm{{\rm x}}_M)\right\}.
\end{equation}

For free-space propagation, the classical indoor scenarios is considered where the user is in line-of-sight (LoS) with the PAs \cite{bereyhi2025mimo}. The free-space channel vector from the $m$-th waveguide to the user can be written as
\begin{equation}\label{sec2A_eq3}
	\bm{{\rm h}}^H_{m}(\bm{{\rm x}}_m)\!\!=\!\!\!\left[\!\frac{\sqrt{\eta}{\rm e}^{-j\frac{2\pi}{\lambda}\Vert \bm{\psi}^u - \bm{\alpha}_{m,1} \Vert}}{\Vert \bm{\psi}^u - \bm{\alpha}_{m,1} \Vert},\!\cdots\!,\!\frac{\sqrt{\eta}{\rm e}^{-j\frac{2\pi}{\lambda}\Vert \bm{\psi}^u - \bm{\alpha}_{m,N} \Vert}}{\Vert \bm{\psi}^u - \bm{\alpha}_{m,N} \Vert}\!\right],
\end{equation}
where $\eta=c^2/(16\pi^2f_c^2)$ is a constant with $c$ and $f_c$ being the speed of light and the carrier frequency, respectively. $\Vert \bm{\psi}^u - \bm{\alpha}_{m,n} \Vert$ represents the distance between the $n$-th PA on the $m$-th waveguide and the user, which is computed by $\Vert \bm{\psi}^u - \bm{\alpha}_{m,n} \Vert = \sqrt{(x^u-x_{m,n}^p)^2+(y^u-y_m)^2+h^2}$. Thus, the overall free-space channel vector $\bm{{\rm h}}^{{\rm H}}(\bm{{\rm X}})\in \mathbb{C}^{1\times MN}$ is given as follows: 
\begin{equation}\label{sec2A_eq4}
	\bm{{\rm h}}^{{\rm H}}(\bm{{\rm X}})=\left[\bm{{\rm h}}_{1}^{{\rm H}}(\bm{{\rm x}}_1), \bm{{\rm h}}_{2}^{{\rm H}}(\bm{{\rm x}}_2), ..., \bm{{\rm h}}_{M}^{{\rm H}}(\bm{{\rm x}}_M)\right].
\end{equation}

Let $s\in\mathbb{C}$ and $\bm{{\rm w}}\in\mathbb{C}^{M\times 1}$ denote the transmitted signal with $\mathbb{E}[|s|^2]=1$ and digital beamforming vector, respectively. Therefore, the received signal at the user can be expressed as
\begin{equation}\label{sec2A_eq5}
	y=\bm{{\rm h}}^{\rm H}(\bm{{\rm X}})\bm{{\rm G}}(\bm{{\rm X}})\bm{{\rm w}}s+z,
\end{equation}
where $z\sim\mathcal{CN}(0,\sigma^2)$ denotes the additive white Gaussian noise (AWGN) with a power of $\sigma^2$. Thus, the SNR is given as
\begin{equation}\label{sec2A_eq6}
	\gamma=\frac{\vert \bm{{\rm h}}^{\rm H}(\bm{{\rm X}})\bm{{\rm G}}(\bm{{\rm X}})\bm{{\rm w}} \vert^2}{\sigma^2}.
\end{equation}
According to \cite{zhou2022rate, gan2025dual}, the SE $f_{{\rm SE}}$ and EE $f_{{\rm EE}}$ for the considered scenario is defined as
\begin{equation}\label{sec2A_eq7}
	f_{{\rm SE}}(\bm{{\rm w}},\! \bm{{\rm X}})\!\!=\!\!{\rm log}_2(1+\gamma),
\end{equation}
\begin{equation}\label{sec2A_eq8}
	f_{{\rm EE}}(\bm{{\rm w}}, \!\bm{{\rm X}}) \!\!=\!\! \frac{f_{{\rm SE}}(\bm{{\rm w}},\! \bm{{\rm X}})}{\Vert \bm{{\rm w}} \Vert^2 +P_f+\! \chi f_{{\rm SE}}(\bm{{\rm w}}, \!\bm{{\rm X}})}.
\end{equation}
Here, $\Vert \bm{{\rm w}} \Vert^2$ is the transmit power consumption. $\chi \geq 0$ is a constant demonstrating the coding, decoding, and backhual power consumption per unit data rate (W/(bit/s/Hz)), respectively. $P_f$ denotes the rate-independent power consumption, which can be modeled as\cite{wang2023simultaneously}
\begin{equation}\label{sec2A_eq9}
	P_f=P_{{\rm BS}}+P_{{\rm BB}}+N_{{\rm RF}}P_{{\rm RF}}+P_{{\rm UE}}+MNP_{{\rm PA}},
\end{equation}
with $P_{{\rm BS}}, P_{{\rm BB}}, P_{{\rm RF}}$, and $P_{{\rm UE}}$ being the power consumption of the oscillator and circuit at the BS, the baseband processing, each RF chain and the circuit at each user, respectively. $N_{{\rm RF}}$ is the number of RF chains. $P_{{\rm PA}}=P^{{\rm act}}_{{\rm PA}}+P^{{\rm mot}}_{{\rm PA}}+P^{{\rm pie}}_{{\rm PA}}$ is the total power consumption of each PA, with $P^{{\rm act}}_{{\rm PA}}$, $P^{{\rm mot}}_{{\rm PA}}$, and $P^{{\rm pie}}_{{\rm PA}}$ being the power consumption of the activation, motorized-module, and piezoelectric-module of each PA, respectively. 

\subsection{Problem Formulation}\label{problem_formulation_for_single_user_PASS}
We aim to balance the SE and EE performance of PASS by jointly optimizing the transmit and pinching beamforming. The resulting multi-object optimization (MOO) problem incorporates two conflicting performance metrics, i.e., SE and EE. The weighted product method is adopted to convert the MOO problem into a single-object optimization (SOO) problem as follows: \footnote{In contrast to conventional weighted-sum methods, weighted-product method avoids scalarization of objective functions, which can help simplify the analysis \cite{Obiedollah2025energy}.}
\begin{subequations}
	\begin{align}
		(\textbf{P1-1}): \max_{\bm{{\rm w}}, \bm{{\rm X}}} & \quad [f_{{\rm SE}}(\bm{{\rm w}}, \bm{{\rm X}})]^{\beta} \times [f_{{\rm EE}}(\bm{{\rm w}}, \bm{{\rm X}})]^{1-\beta} \label{sec2B_eq1}\\
		\mathrm{s.t.} & \quad x_{m,n}^p \in [-\frac{D_x}{2}, \frac{D_x}{2}], \forall m, \forall n, \label{sec2B_eq2}\\
		& \quad \vert x_{m,n}^p - x_{m,n-1}^p\vert \geq \Delta_{{\rm min}}, \forall m, \forall n, \label{sec2B_eq3}\\
		& \quad \Vert \bm{{\rm w}} \Vert^2 \leq P_{{\rm T}}, \label{sec2B_eq4}
	\end{align}
\end{subequations}
where constraint (\ref{sec2B_eq2}) ensures that each PA is positioned within the valid range of the waveguide, $\Delta_{{\rm min}}$ in (\ref{sec2B_eq3}) is the minimum spacing required to prevent mutual coupling between the PAs \cite{zhaolin2025modeling}, and (\ref{sec2B_eq4}) refers to the constraint on transmit power at the BS side. It should be noted that the problem reduces to a conventional SE-maximization problem when $\beta=1$, and to an EE-maximization problem when $\beta=0$. By tuning the preference weight $\beta$, this framework provides additional freedom to strike a balance between the conflicting objectives. Moreover, the weighted product method can provide a Pareto-optimal solution to the original MOO problem, as demonstrated in Theorem \ref{theorem1}.
\begin{theorem}
	\label{theorem1}
	The weighted product method yields a Pareto-optimal solution $(\bm{{\rm w}}^*, \bm{{\rm X}}^*)$ to the SE-EE MOO problem.
\end{theorem}
\begin{IEEEproof}
	Please refer to Appendix \ref{proof_of_theorem1}.
\end{IEEEproof}

Noted that problem (\textbf{P1-1}) is non-convex since $\bm{{\rm w}}$ and $\bm{{\rm X}}$ are strongly coupled in (\ref{sec2B_eq1}). Fortunately, for any given PA positions $\bm{{\rm X}}$, the optimal transmit beamforming $\bm{{\rm w}}^*$ of \textbf{(P1-1)} can be obtained by using MRT strategy:
\begin{equation}\label{sec2B_eq5}
	\bm{{\rm w}}^* = \sqrt{P}\frac{(\bm{{\rm h}}^{{\rm H}}(\bm{{\rm X}})\bm{{\rm G}}(\bm{{\rm X}}))^H}{\Vert \bm{{\rm h}}^{{\rm H}}(\bm{{\rm X}})\bm{{\rm G}}(\bm{{\rm X}})\bm{{\rm w}} \Vert},
\end{equation}
where $P$ is the scalar transmit power. Therefore, the SNR for the user under MRT strategy can be given by $\gamma=P\Vert \bm{{\rm h}}^{{\rm H}}(\bm{{\rm X}})\bm{{\rm G}}(\bm{{\rm X}}) \Vert^2/\sigma^2$. Accordingly, the SE and EE can be rewritten as
\begin{equation}\label{sec2B_eq6}
	f_{{\rm SE}}(P, \bm{{\rm X}})={\rm log}_2(1+\gamma),
\end{equation}
\begin{equation}\label{sec2B_eq7}
	f_{{\rm EE}}(P, \bm{{\rm X}}) = \frac{f_{{\rm SE}}(P, \bm{{\rm X}})}{P + P_f + \chi f_{{\rm SE}}(P, \bm{{\rm X}})}.
\end{equation}
Based on the above transformation, the optimization of the digital beamforming vector reduces to the design of the scalar transmit power $P$. In the following, we first optimize one variable by fixing the other, and then a two-stage optimization framework is demonstrated.

\subsection{Optimization of PA Positions}\label{sec:2:pa_placement}
To facilitate the optimization of PA positions, we first transform problem (\textbf{P1-1}) into the following equivalent form for a given $P$:
\begin{equation}\label{sec2C_eq1}
	(\textbf{P1-2}): \max_{\bm{{\rm X}}} \; \beta {\rm ln} f_{{\rm SE}}(\bm{{\rm X}})\!+\!(1-\beta){\rm ln}f_{{\rm EE}}(\bm{{\rm X}}) \; \mathrm{s.t.} \; (\ref{sec2B_eq2}), (\ref{sec2B_eq3}) 
\end{equation}
\begin{lemma}
	\label{lemma1}
	The objective function of problem (\textbf{P1-2}) is strictly increasing with $f_{{\rm SE}}(\bm{{\rm X}})$.
\end{lemma}

\begin{IEEEproof}
	Let $f$ denote the objective function in (\textbf{P1-2}) and take the first derivative of $f$ with respect to $f_{{\rm SE}}(\bm{{\rm X}})$ leads to
	\begin{equation}\label{sec2C_eq2}
		f'\!=\!\frac{\beta}{f_{{\rm SE}}(\bm{{\rm X}})}\!+\!\frac{1-\beta}{f_{{\rm EE}}(\bm{{\rm X}})}\frac{P+P_f}{(P+P_f+\chi f_{{\rm SE}}(\bm{{\rm X}}))^2},
	\end{equation}
	it is clear that $f'>0$, so $f$ is increasing with $f_{{\rm SE}}(\bm{{\rm X}})$. The proof is thus completed.
\end{IEEEproof}
\noindent Based on \textbf{Lemma} \textbf{\ref{lemma1}}, problem (\textbf{P1-2}) is equivalent to a SE-maximization problem, which can be formulated as
\begin{equation}\label{sec2C_eq3}
	(\textbf{P1-3}): \max_{\bm{{\rm X}}} \; {\rm log}_2\left(1+\gamma\right) \;
	\mathrm{s.t.} \; (\ref{sec2B_eq2}), (\ref{sec2B_eq3}).
\end{equation}
From problem (\textbf{P1-3}), it is noticed that maximizing the SE is equivalent to maximizing the SNR. Therefore, problem (\textbf{P1-3}) can be equivalently recast as follows:
\begin{subequations}
	\begin{align}
		(\textbf{P1-4}): \max_{\textbf{X}} & \; \left| \sum_{m=1}^M\sum_{n=1}^N \frac{{\rm e}^{-j\phi_{m,n}}}{\Vert \bm{\alpha}_{m,n}-\bm{\psi}^u \Vert \times {\rm e}^{a\Vert \bm{\alpha}_{m,0} - \bm{\alpha}_{m,n} \Vert}} \right| \label{sec2C_eq4}\\
		\mathrm{s.t.} & \; (\ref{sec2B_eq2}), (\ref{sec2B_eq3}), \label{sec2C_eq5} \\
		& \; \phi_{m,n} \!\!=\!\! \frac{2\pi}{\lambda}\!\Vert \bm{\alpha}_{m,n}\!\!-\!\!\bm{\psi}^u \Vert\!\!+\!\!\frac{2\pi}{\lambda_g}\Vert \bm{\alpha}_{m,0} \!-\!\! \bm{\alpha}_{m,n} \!\Vert. \label{sec2C_eq6}
	\end{align}
\end{subequations}
The objective function in (\textbf{P1-4}) reveals that the placement of PAs affects both attenuation (large-scale path loss and in-waveguide attenuation) and the phase shifts. Following the two-stage optimization framework presented in \cite{xu2025rate}, we can obtain a coarse PA deployment without considering phase shifts, followed by a wavelength-scale adjustment of the PA positions to align the phases. For coarse PA position optimization, the objective function of problem (\textbf{P1-4}) can be decomposed into $MN$ sub-objective functions, each corresponding to an optimal PA position. This allows us to sequentially optimize each PA position to maximize the overall objective function. Following the analysis presented in \cite{xu2025pinching}, for the $n$-th PA on the $m$-th waveguide, its corresponded sub-objective function can be given by
\begin{equation}\label{sec2C_eq7}
	f_{m,n}^{\rm sub}=\left[(x_{m,n}^p\!-\!x^u)^2+(y_m-y^u)^2+h^2\right]^{-\frac{1}{2}}{\rm e}^{-a(x_{m,n}^p\!+\!\frac{D_x}{2})}.
\end{equation}
It is noticed that $f_{m,n}^{\rm sub}\geq 0$ for any feasible $x_{m,n}^p$. Let $D_m=(y_m-y^u)^2+h^2$, therefore, it is equivalent to finding $x_{m,n}^p$ which minimizes
\begin{equation}\label{sec2C_eq8}
	f_{m,n}^{\rm sub, 2}(x_{m,n}^p)\!=\!(x_{m,n}^p-x^u)^2{\rm e}^{2a(x_{m,n}^p+\frac{D_x}{2})}+D_m{\rm e}^{2a(x_{m,n}^p+\frac{D_x}{2})}.
\end{equation}
To find $x_{m,n}^p$ that minimizes $f_{m,n}^{\rm sub, 2}(x_{m,n}^p)$, we first calculate its first-order derivative, which is given by
\begin{align}
	[f_{m,n}^{\rm sub, 2}(x_{m,n}^p)]'&= {\rm e}^{2a(x_{m,n}^p+\frac{D_x}{2})}\times \nonumber\\
	&\left[2(x_{m,n}^p-x^u)+2a(x_{m,n}^p-x^u)^2+2aD_m\right]. \label{sec2C_eq9}
\end{align}
Since ${\rm e}^{2a(x_{m,n}^p+\frac{D_x}{2})} > 0$, we only need to focus on the following function:
\begin{equation}\label{sec2C_eq10}
	g_{m,n}(x_{m,n}^p) \triangleq a(x_{m,n}^p-x^u)^2+(x_{m,n}^p-x^u)+aD_m.
\end{equation}
Further, rewrite the above equation in quadratic form yields
\begin{equation}\label{sec2C_eq11}
	g_{m,n}(x_{m,n}^p) = a\left(x_{m,n}^p-x^u+\frac{1}{2a}\right)^2+aD_m-\frac{1}{4a}
\end{equation}
\textbf{Case 1}: If ${D_m \geq \frac{1}{4a^2}}$, $g_{m,n}(x_{m,n}^p)\geq 0$ always holds, which implies that function $f_{m,n}^{\rm sub, 2}(x)$ is monotonically increasing. Thus, the optimal position of the $n$-th PA on the $m$-th waveguide can be obtained as\footnote{It should be noted that for the PAs on the same waveguide, $D_m$ remains unchanged, so the monotonicity of function $f_{m,n}^{\rm sub, 2}(x)$ would not change.}
\begin{equation}\label{sec2C_eq12}
	x_{m,n}^p = -\frac{D_x}{2}+(n-1)\Delta_{{\rm min}}, n=1,2,\cdots,N.
\end{equation}
\textbf{Case 2}: If ${D_m < \frac{1}{4a^2}}$, the quadratic equation $g_{m,n}(x_{m,n}^p)=0$ has two real roots, which can be given by
\begin{equation}\label{sec2C_eq13}
	x_{m,n}^{p,1} = x^u + \frac{-1-\sqrt{1-4a^2D_m}}{2a},
\end{equation}
\begin{equation}\label{sec2C_eq14}
	x_{m,n}^{p,2} = x^u + \frac{-1+\sqrt{1-4a^2D_m}}{2a},
\end{equation}
where $x_{m,n}^{p,1} < x_{m,n}^{p,2} < x^u$ holds. Consequently, there are three sub-cases needed to be considered:

\textbf{Case 2.1}: $x_{m,n}^{p,1} < x_{m,n}^{p,2} \leq -\frac{D_x}{2}$. In this case, the function $f_{m,n}^{\rm sub, 2}(x)$ is monotonically increasing for $x\in [-\frac{D_x}{2}, \frac{D_x}{2}]$. Thus, for all $N$ PAs on this waveguide, their coarse positions can be obtained as in (\ref{sec2C_eq12}).   

\begin{figure}[t]
	\centering
	\includegraphics[width=3.3in]{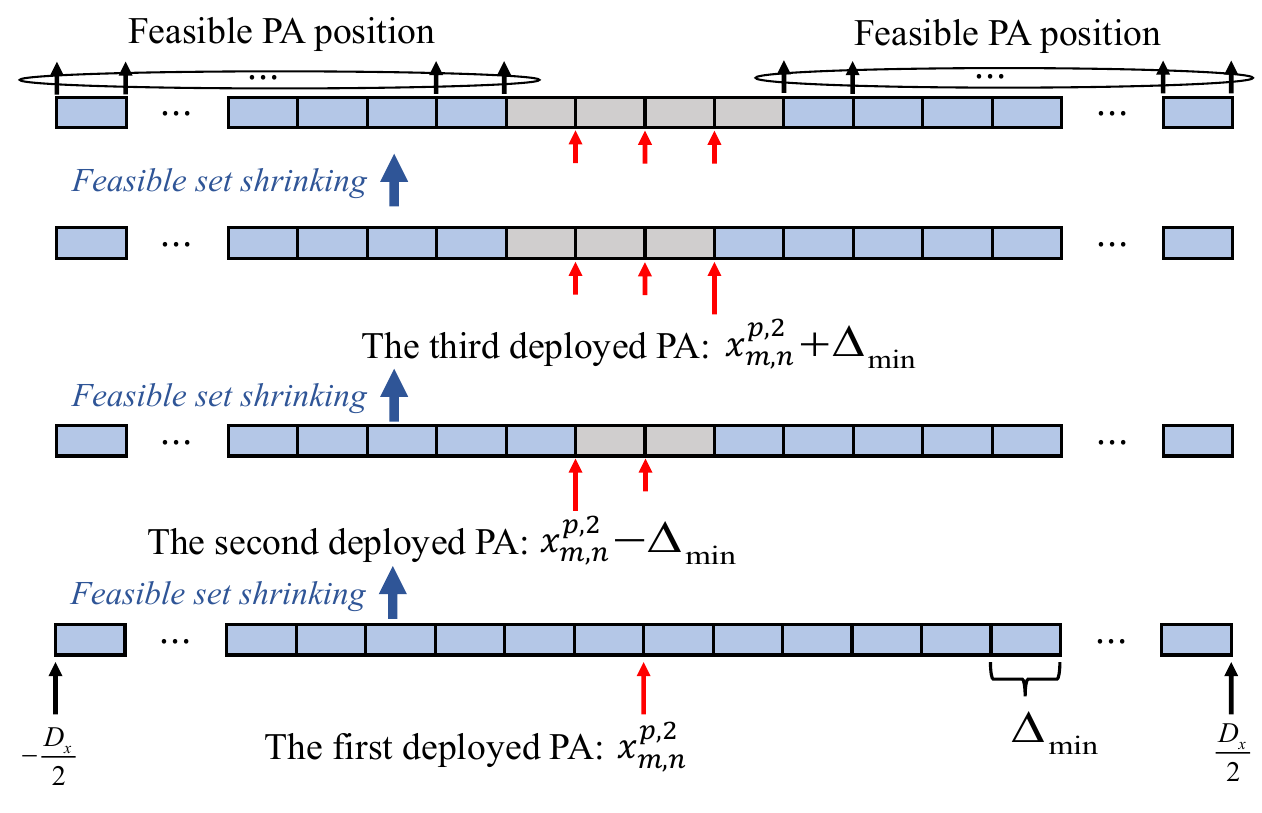}
	\caption{The illustration of feasible set shrinking in Case 2.2.}
	\label{figure_2}
\end{figure}

\textbf{Case 2.2}: $x_{m,n}^{p,1} \leq -\frac{D_x}{2} < x_{m,n}^{p,2}$. In this case, function $f_{m,n}^{\rm sub, 2}(x)$ is monotonically decreasing for $x\in [-\frac{D_x}{2}, x_{m,n}^{p,2})$, and monotonically increasing for $x\in [x_{m,n}^{p,2}, \frac{D_x}{2}]$. Thus, the optimal position of this specific PA is $x_{m,n}^p=x_{m,n}^{p,2}$. For the next PA to be deployed, we need to compare $f_{m,n}^{\rm sub, 2}(x_{m,n}^{p,2}-\Delta_{{\rm min}})$ and $f_{m,n}^{\rm sub, 2}(x_{m,n}^{p,2}+\Delta_{{\rm min}})$ to determine the position. For example, if the next PA is deployed at $x_{m,n}^{p,2}-\Delta_{{\rm min}}$, then the feasible set for the subsequent PA becomes $[-\frac{D_x}{2}, x_{m,n}^{p,2}-2\Delta_{{\rm min}}]\cup [x_{m,n}^{p,2}+\Delta_{{\rm min}}, \frac{D_x}{2}]$. The illustration of this feasible set changing is shown in Fig. \ref{figure_2}.

\begin{figure}[t]
	\centering
	\includegraphics[width=3.3in]{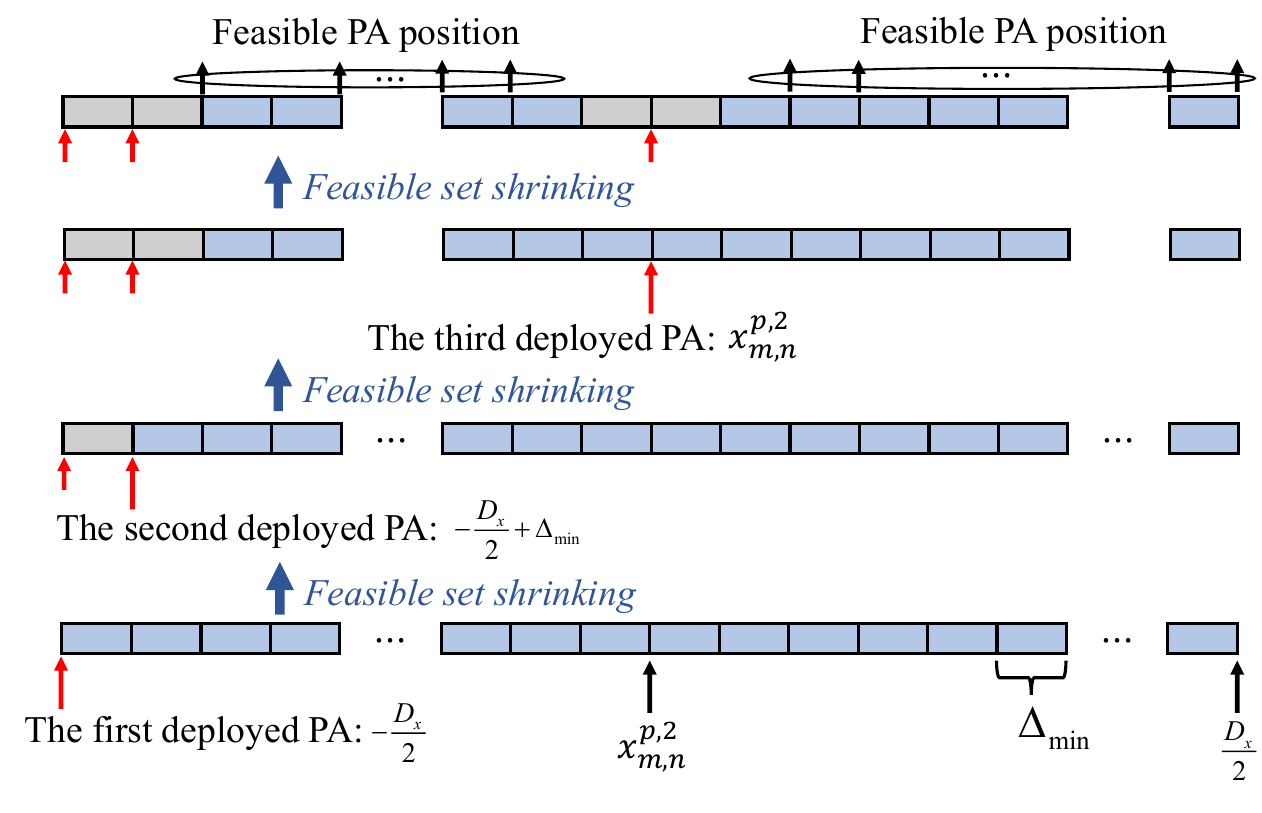}
	\caption{The illustration of feasible set shrinking in Case 2.3.}
	\label{figure_3}
\end{figure}

\textbf{Case 2.3}: $-\frac{D_x}{2} < x_{m,n}^{p,1} < x_{m,n}^{p,2}$. In this case, function $f_{m,n}^{\rm sub, 2}(x)$ is monotonically increasing for $x\in [-\frac{D_x}{2}, x_{m,n}^{p,1}) \cup [x_{m,n}^{p,2}, \frac{D_x}{2}]$, and monotonically decreasing for $x\in [x_{m,n}^{p,1}, x_{m,n}^{p,2})$. Therefore, for the first deployed PA, we need to compare $f_{m,n}^{\rm sub, 2}(-\frac{D_x}{2})$ and $f_{m,n}^{\rm sub, 2}(x_{m,n}^{p,2})$. For instance, if $f_{m,n}^{\rm sub, 2}(-\frac{D_x}{2})$ is smaller, then this PA will be deployed at $-\frac{D_x}{2}$, and for the next PA, we need to compare $f_{m,n}^{\rm sub, 2}(-\frac{D_x}{2}+\Delta_{{\rm min}})$ and $f_{m,n}^{\rm sub, 2}(x_{m,n}^{p,2})$. Subsequently, if the next PA is deployed at $x_{m,n}^{p,2}$, the feasible set becomes $[-\frac{D_x}{2}+\Delta_{{\rm min}}, x_{m,n}^{p,2}-\Delta_{{\rm min}}]\cup [x_{m,n}^{p,2}+\Delta_{{\rm min}}, \frac{D_x}{2}]$. The illustration of this feasible set changing is shown in Fig. \ref{figure_3}.

An important observation from the above analysis is that \textit{after coarse deployment, the PAs will be divided into one or two clusters, where the spacing between adjacent PAs within each cluster is $\Delta_{{\rm min}}$}, as illustrated in the gray areas in Fig. \ref{figure_2} and \ref{figure_3}. Based on these features, we propose a new refinement framework to ensure constructive signal combination at the user\footnote{It should be noted that the two-stage PA deployment framework essentially assumes that the refinement of the position at the wavelength scale does not significantly affect the free-space path loss and the in-waveguide attenuation.}. In contrast to the refinement method in \cite{xu2025rate} that relies on the coarse position of the previous PA to determine the searching space for the current PA, which may cause potential overlap in search spaces and subsequently, violate the constraint (\ref{sec2B_eq3}). To address this issue while generalizing the PA refinement to more general multi-waveguide case, this paper proposes a new sequential refinement framework for multi-waveguide PASS. Furthermore, we also provide an analysis of the worst-case deviation of a PA from its initial coarse position, which can offer practical engineering insights into determining the number of PAs to deploy. We will introduce the refinement procedure separately in two cases below.

\subsubsection{Refine on the Same Waveguide}

We first consider refining the PAs located on the same waveguide. The position of the first PA ($x_{1,1}^{p,{\rm opt}}$) on the first waveguide is taken as the ``reference point''\footnote{Selecting $x_{1,1}^{p,{\rm opt}}$ as the reference point is to mitigate the propagation of cumulative errors.} for phase alignment. If the $n$-th PA (current PA) and the previous one belong to the same cluster, as shown in Fig. \ref{figure_2}, then the searching space of the current $n$-th PA is $[x_{m,n-1}^{p,{\rm opt}}+\Delta_{{\rm min}}, x_{m,n-1}^{p,{\rm opt}}+3\Delta_{{\rm min}}]$. If they do not belong to the same cluster, as shown by the second and third PAs in Fig. \ref{figure_2}, then the searching space of the current $n$-th PA is $[{\rm max}\left(x_{m,n-1}^{p,{\rm opt}}+\Delta_{{\rm min}}, x_{m,n}^{p}-2\Delta_{{\rm min}}\right), x_{m,n}^{p}]$. After determining the one-dimensional searching space, the refined position of the $n$-th PA is obtained by using the first-found position that minimizes ${\rm mod}\left\{\phi_{1,1}-\phi_{m,n}, 2\pi\right\}$, where ${\rm mod}\{a,b\}$ represents the module operation of $a$ by $b$. 

\subsubsection{Refine Across Different Waveguides}

When refining on the first waveguide, the position of the first PA on this waveguide can be directly used as the reference point without adjustment. However, for the $m$-th waveguide ($m>1$), the position of its first PA requires refinement. Here, we describe the refinement of the first PA on this $m$-th waveguide; adjustments to the remaining PAs on this waveguide follow the method introduced in the previous part. Specifically, let $x_{1,1}^{p,{\rm opt}}$ as the reference point, the searching space of the first PA on the $m$-th waveguide is $[{\rm max}\left(-\frac{D_x}{2}, x_{m,1}^{p}-2\Delta_{{\rm min}}\right), x_{m,1}^{p}]$. Next, the refined position of this PA is obtained by using the first-found position that minimizes ${\rm mod}\left\{\phi_{1,1}-\phi_{m,1}, 2\pi\right\}$.
\begin{algorithm}[t]
	\caption{The proposed PA placement framework}
	\label{algorithm1}
	\begin{algorithmic}
		\STATE \textbf{Initialize}: The user location $\bm{\psi}^u$, $M$, $N$, $h$, $D_x, D_y$, $\Delta_{{\rm min}}$, and the attenuation coefficient $a$.
		\STATE Get the rough position of PAs based on Case 1 and Case 2 described in Section \ref{sec:2:pa_placement}.
		\STATE Get the phase shift corresponded to $x_{1,1}^{p,{\rm opt}}$.
		\FOR{$m=1$ to $M$}
		\STATE /*\emph{Refine on the same waveguide} */
		\FOR{$n=2$ to $N$} 
		\STATE Obtain the searching space for the $n$-th PA.
		\STATE Conduct one-dimensional search to get $x_{m,n}^{p,{\rm opt}}$.
		\ENDFOR
		\STATE /*\emph{Refine across different waveguide} */
		\IF{$m<M$}
		\STATE Obtain the searching space for the first PA on the $m+1$-th waveguide.
		\STATE Conduct one-dimensional search to get $x_{m+1,1}^{p,{\rm opt}}$.
		\ENDIF
		\ENDFOR
	\end{algorithmic}
\end{algorithm}

\begin{figure}[t]
	\centering
	\includegraphics[width=3.4in]{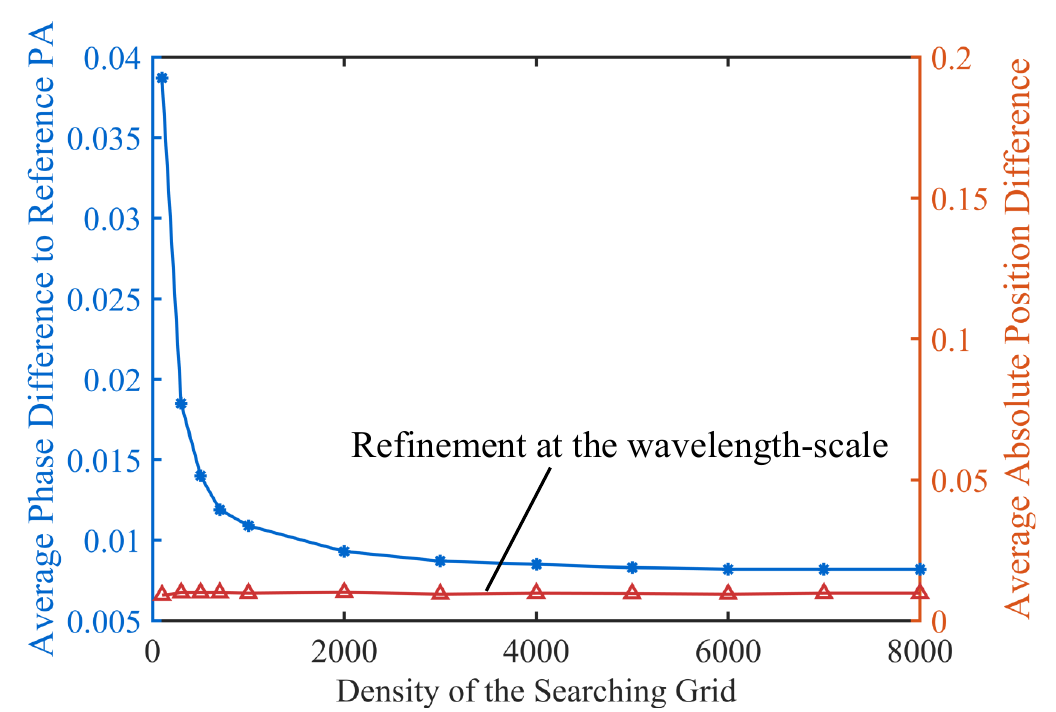}
	\caption{Convergence quality and distance offset of the refinement algorithm.}
	\label{figure_4}
\end{figure}

By refining each PA on the first through the $M$-th waveguide sequentially, we can obtain the final positions of PASS. The overall PA placement algorithm is summarized in \textbf{Algorithm} \textbf{\ref{algorithm1}}. It is noticed that the grid search naturally supports discrete PA activation. As the grid density increases, the average phase difference between the PAs and the reference point, as well as the average distance between the refined and coarse PA positions, are shown in Fig. \ref{figure_4}. It can be seen that our algorithm achieves an acceptable phase alignment error when the grid density exceeds 3000. Furthermore, the refinement of PA positions can be confined to the wavelength scale. 
\begin{remark}
	In the worst case, the maximum absolute distance between a PA's refined position on the waveguide and its original coarse position is $(2N-2)\Delta_{{\rm min}}$. Given that the number of PAs per waveguide is typically small under single-user scenarios, our algorithm guarantees that the refinement is confined to the wavelength-scale. Therefore, it is safe to neglect the in-waveguide attenuation in PA refinement stage.
\end{remark}

\subsection{Optimization of Transmit Power}\label{sec:2:power_optimization}
In this subsection, we focus on the optimization of transmit power $P$ at BS given fixed PA positions $\bm{{\rm X}}$. By defining $\zeta = \Vert \bm{{\rm h}}^{{\rm H}}(\bm{{\rm X}})\bm{{\rm G}}(\bm{{\rm X}}) \Vert^2/\sigma^2$, the SE and EE expressions can be simplified into
\begin{equation}\label{sec2D_eq1}
	f_{{\rm SE}}(P)={\rm log}_2\left(1+\zeta P\right),
\end{equation}
\begin{equation}\label{sec2D_eq2}
	f_{{\rm EE}}(P) \!=\! \frac{{\rm log}_2\left(1+\zeta P\right)}{P \!+\! P_f \!+\! \chi {\rm log}_2\left(1+\zeta P\right)}.
\end{equation}
Similarly, by taking the logarithm of the objective function (\ref{sec2B_eq1}), the subproblem with respect to the transmit power $P$ is given by
\begin{subequations}
	\setlength\abovedisplayskip{3pt}
	\setlength\belowdisplayskip{3pt}
	\begin{equation}
		\setlength\abovedisplayskip{3pt}
		\setlength\belowdisplayskip{3pt}
		(\textbf{P1-5}): \max_{P} \!\; \beta {\rm ln} f_{{\rm SE}}(P)\!+\!\!(1\!-\!\beta) {\rm ln} f_{{\rm EE}}(P) \;\mathrm{s.t.} \; 0\leq P \leq P_T.\tag{27}
	\end{equation}
\end{subequations}
To determine the optimal transmit power, it is essential to analyze the properties of $f_{{\rm SE}}(P)$ and $f_{{\rm EE}}(P)$. Thus, we present the following lemma and proposition.
\begin{lemma}
	\label{lemma2}
	There exists one and only one point $P^*\in [0, +\infty)$ that maximizes $f_{{\rm EE}}(P)$, and $f_{{\rm EE}}(P)$ is strictly increasing and concave at $[0, P^*]$ while strictly decreasing and only quasi-concave at $(P^*, +\infty)$.
\end{lemma}
\begin{IEEEproof}
	Please refer to Appendix \ref{proof_of_lemma2}.
\end{IEEEproof}
\begin{proposition}
	\label{proposition1}
	The Pareto optimal set of problem (\textbf{P1-5}) is:
	\begin{equation}
		\setlength\abovedisplayskip{3pt}
		\setlength\belowdisplayskip{3pt}
		\label{sec2D_eq4}
		\mathcal{P}=\left\{
		\begin{aligned}
			&\{P | P=P_T\} \quad &\text{if} \quad P^* \geq P_T, \\
			&\{P | P^* \leq P \leq P_T\} \quad &\text{if} \quad P^* < P_T.
		\end{aligned}
		\right.
	\end{equation}
\end{proposition} 
\begin{IEEEproof}
	Please refer to Appendix \ref{proof_of_proposition1}.
\end{IEEEproof}

From \textbf{Proposition} \textbf{\ref{proposition1}}, in the case of $P^* \geq P_T$, the Pareto optimal set $\mathcal{P}$ contains a single point, i.e., $P_T$, which means that the globally optimal solution for problem (\textbf{P1-5}) is unique. Therefore, we will then focus on the case when $P^* < P_T$. Let $f_2(P)$ denote the objective function in problem (\textbf{P1-5}), we have
\begin{equation}\label{sec2D_eq5}
	f_2(P)={\rm ln} f_{{\rm SE}}(P)-(1-\beta){\rm ln}(P+P_f+\chi f_{{\rm SE}}(P)).
\end{equation}

\begin{figure*}[b]
	\centering
	\hrulefill
	\begin{align}
		&f_2'(P)=\frac{\frac{\zeta}{1+\zeta P}}{{\rm ln}(1+\zeta P)}-(1-\beta)\frac{1+\frac{\chi}{{\rm ln}2}\frac{\zeta}{1+\zeta P}}{P+P_f+\frac{\chi}{{\rm ln}2}{\rm ln}(1+\zeta P)} =\frac{\frac{\zeta \left(P+P_f+\frac{\chi}{{\rm ln}2}{\rm ln}(1+\zeta P)\right)}{\left(1+\zeta P + \frac{\chi \zeta}{{\rm ln}2}\right){\rm ln}(1+\zeta P)} - (1-\beta)}{(1+\zeta P){\rm ln}(1+\zeta P)\left(P+P_f+\frac{\chi}{{\rm ln}2}{\rm ln}(1+\zeta P)\right)} \label{sec2D_eq6} \\
		& g_2(P)=\frac{\zeta P}{\left(1+\zeta P + \frac{\chi \zeta}{{\rm ln}2}\right){\rm ln}(1+\zeta P)} +\frac{\zeta P_f}{\left(1+\zeta P + \frac{\chi \zeta}{{\rm ln}2}\right){\rm ln}(1+\zeta P)} + \frac{\frac{\zeta \chi}{{\rm ln}2}}{1+\zeta P+\frac{\zeta \chi}{{\rm ln}2}} \label{sec2D_eq7}\\
		& \left[\frac{\zeta P}{\left(1+\zeta P + A\right){\rm ln}(1+\zeta P)}\right]'=\frac{\zeta {\rm ln}(1+\zeta P)+A\zeta {\rm ln}(1+\zeta P)-\zeta^2 P-\frac{\zeta^2 AP}{1+\zeta P}}{[(1+\zeta P+A){\rm ln}(1+\zeta P)]^2} \label{sec2D_eq8}
	\end{align}
\end{figure*}
\noindent To maximize $f_2(P)$, we derive its first derivative, which is given by (\ref{sec2D_eq6}) at the bottom of next page. Let $g_2(P)$ denotes the first term in the numerator of (\ref{sec2D_eq6}), i.e.,
\begin{equation}\label{sec2D_eq9}
	\setlength\abovedisplayskip{3pt}
	\setlength\belowdisplayskip{3pt}
	g_2(P)=\frac{\zeta \left(P+P_f+\frac{\chi}{{\rm ln}2}{\rm ln}(1+\zeta P)\right)}{\left(1+\zeta P + \frac{\chi \zeta}{{\rm ln}2}\right){\rm ln}(1+\zeta P)}.
\end{equation}
Thus, the numerator of $f'_2(P)$ can be expressed as $g_2(P)-(1-\beta)$. To analyze the roots of the equation $f'_2(P)=0$, we first need to analyze the monotonicity of function $g_2(P)$, which leads to the following lemma.
\begin{lemma}
	\label{lemma3}
	$g_2(P)$ is strictly decreasing with $P$.
\end{lemma}
\begin{IEEEproof}
	Please refer to Appendix \ref{proof_of_lemma3}.
\end{IEEEproof}

\noindent Setting $f_2'(P)=0$ yields $g_2(P) = 1-\beta$. As we previously demonstrated in \textbf{Lemma} \textbf{\ref{lemma3}}, $g_2(P)$ is strictly decreasing with $P$, leading to $0<g_2(P_T)\leq g_2(P) \leq g_2(P^*)$. Furthermore, recall that $P^*$ is the maximum point of $f_{{\rm EE}}(P)$, that is, $P^*$ satisfies $\zeta(P^*+P_f)=(1+\zeta P^*){\rm ln}(1+\zeta P^*)$ based on (\ref{proof_lemma2_eq1}). Therefore, we will have
\begin{align}
	g_2(P^*) &= \frac{\zeta \left(P^*+P_f+\frac{\chi}{{\rm ln}2}{\rm ln}(1+\zeta P^*)\right)}{\left(1+\zeta P^* + \frac{\chi \zeta}{{\rm ln}2}\right){\rm ln}(1+\zeta P^*)} \nonumber \\
	&=\frac{\zeta (P^*+P_f)+\frac{\zeta \chi}{{\rm ln}2}{\rm ln}(1+\zeta P^*)}{\zeta (P^*+P_f)+\frac{\zeta \chi}{{\rm ln}2}{\rm ln}(1+\zeta P^*)} =1. \label{sec2D_eq10}
\end{align}
From (\ref{sec2D_eq10}) we can know that $g_2(P)\in (0, 1]$, since $1-\beta\in [0,1]$, thus, we need to consider the following two cases:

\textbf{Case 1}: $g_2(P_T) > 1-\beta$. In this case, it is evident that $f_2'(P)>0$, which means that $f_2(P)$ is strictly increasing with $P$. Therefore, the optimal transmit power is $P_T$.
	
\textbf{Case 2}: $g_2(P_T) \leq 1-\beta$. In this case, we can know that the equation $g_2(P)=1-\beta$ has a unique solution, which is denoted as $P^{**}$. Therefore, we can infer that when $P\in [P^*, P^{**}]$, $f_2'(P)\geq0$, which means that $f_2(P)$ is increasing in this interval. When $P\in (P^{**}, P_T]$, $f_2'(P)<0$, indicating that $f_2(P)$ is decreasing in this interval. The optimal transmit power is $P^{**}$ in this case.

Therefore, the optimal solution for problem (\textbf{P1-5}) can be given as
\begin{equation}
	\label{sec2D_eq11}
	{\rm P}_{{\rm opt}}=\left\{
	\begin{aligned}
		& P_T, \: \text{if} \: P_T \leq P^* \\
		& P_T, \: \text{if} \: P_T > P^* \: \text{and} \: g_2(P_T) > 1-\beta \\
		& P^{**},  \text{if} \: P_T > P^* \: \text{and} \: g_2(P_T) \leq 1-\beta \\
	\end{aligned}
	\right.
\end{equation}
\begin{remark}
	\label{remark2}
	Based on (\ref{sec2D_eq11}), the optimal transmit power ${\rm P}_{{\rm opt}}$ is controlled by 1) the transmit power limit $P_T$ at the BS; 2) the SE-EE weighting factor $\beta$; and 3) the PA positions ${\rm \textbf{X}}$. When $P_T\leq P^*$, both SE and EE increase monotonically with transmit power for all $\beta\in [0,1]$. Conversely, when $P_T > P^*$, SE and EE become conflicting objectives. These findings necessitates studying the SE-EE trade-off in the regime where $P_T > P^*$, wherein the optimal transmit power achieving the best trade-off is related with $\beta$ and the configuration of PASS.
\end{remark}

\subsection{Overall Algorithm}
As analyzed in subsection \ref{sec:2:pa_placement}, the optimal $\bm{{\rm X}}$ in (\textbf{P1-1}) is independent of $P$ when the constraints in PA placement optimization problem pertain solely to PA positions (i.e., (\ref{sec2B_eq2}) and (\ref{sec2B_eq3})), whereas the optimal transmit power $P$ depends on $\bm{{\rm X}}$ (see (\ref{sec2D_eq11})). Consequently, we can first optimize $\bm{{\rm X}}$ under an arbitrarily given $P$, and then derive the optimal transmit power based on this optimized $\bm{{\rm X}}$. The exact processes are summarized in \textbf{Algorithm} \textbf{\ref{algorithm2}}. 

\begin{algorithm}[t]
	\caption{Two-Stage joint beamforming design}
	\label{algorithm2}
	\begin{algorithmic}[1]
		\STATE \textbf{Initialize}: The user location $\bm{\psi}^u$.
		\STATE Obtain the optimal positions of PA using \textbf{Algorithm} \textbf{\ref{algorithm1}}.
		\STATE Obtain the optimal transmit power based on eq. (\ref{sec2D_eq11}).
	\end{algorithmic}
\end{algorithm}

\section{Joint SE-EE Design For Multi-User PASS}\label{sec:3}
\subsection{System Model}
\begin{figure}[t]
	\centering
	\includegraphics[width=3.5in]{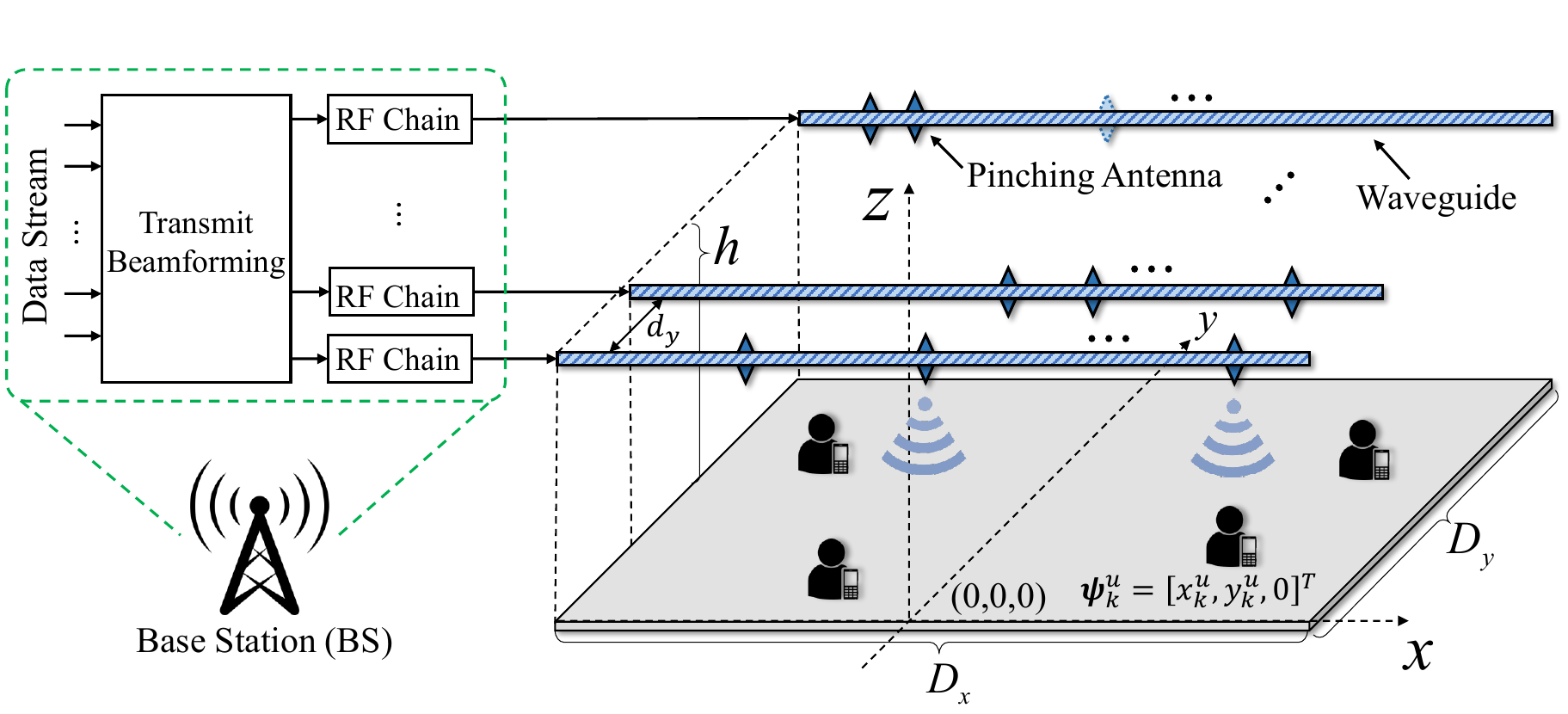}
	\caption{System model for PASS-enabled multi-user communications.}
	\label{figure_5}
\end{figure}

In addition to single-user scenario, the potential of PASS can be further exploited by serving multiple users simultaneously\cite{zhao2025pinching}, as illustrated in Fig. \ref{figure_5}. Let $\mathcal{K}=\{1,2,...,K\}$ denote the set of $K$ single-antenna users, and the position of each user $k\in\mathcal{K}$ is specified as $\psi_k^u=[x_k^u, y_k^u, 0]^{\rm T}$. Let $\bm{{\rm s}}=[s_1,s_2,...,s_K]^{\rm T}\in \mathbb{C}^{K\times1}$ denote the vector of transmit signals, where $s_k\in \mathbb{C}$ is the encoded signal for user $k$. The signals for $K$ users are multiplexed at the baseband using digital transmit beamforming matrix $\bm{{\rm W}}=[\bm{{\rm w}}_1,\bm{{\rm w}}_2,...,\bm{{\rm w}}_K]\in\mathbb{C}^{M\times K}$, where $\bm{{\rm w}}_k$ denotes the transmit beamforming vector for user $k$. Similar to the single-user scenario, the free-space channel vector between the $m$-th waveguide and the $k$-th user can be written as
\begin{equation}\label{sec3A_eq1}
	\bm{{\rm h}}^{\rm H}_{m,k}(\bm{{\rm x}}_m)\!\!=\!\!\!\left[\!\frac{\sqrt{\eta}{\rm e}^{-j\frac{2\pi}{\lambda}\Vert \bm{\psi}_k^u - \bm{\alpha}_{m,1} \Vert}}{\Vert \bm{\psi}_k^u - \bm{\alpha}_{m,1} \Vert},\!\cdots\!,\!\frac{\sqrt{\eta}{\rm e}^{-j\frac{2\pi}{\lambda}\Vert \bm{\psi}_k^u - \bm{\alpha}_{m,N} \Vert}}{\Vert \bm{\psi}_k^u - \bm{\alpha}_{m,N} \Vert}\!\right],
\end{equation}
and the overall free-space channel vector $\bm{{\rm h}}_k^{{\rm H}}(\bm{{\rm X}})\in \mathbb{C}^{1\times MN}$ to user $k$ can be written as
\begin{equation}\label{sec3A_eq2}
	\bm{{\rm h}}_k^{{\rm H}}(\bm{{\rm X}}) = \left[\bm{{\rm h}}_{1,k}^{{\rm H}}(\bm{{\rm x}}_1), \bm{{\rm h}}_{2,k}^{{\rm H}}(\bm{{\rm x}}_2), ..., \bm{{\rm h}}_{M,k}^{{\rm H}}(\bm{{\rm x}}_M)\right].
\end{equation}
The received signal at user $k$ is given by
\begin{equation}\label{sec3A_eq3}
	y_k\!=\!\bm{{\rm h}}_k^{{\rm H}}(\bm{{\rm X}})\bm{G}(\bm{{\rm X}})\bm{{\rm w}}_ks_k\!+\!\!\!\!\!\!\sum_{i=1,i\neq k}^K\bm{{\rm h}}_k^{{\rm H}}(\bm{{\rm X}})\bm{G}(\bm{{\rm X}})\bm{{\rm w}}_is_i \!+ n_k,
\end{equation}
where $n_k \sim \mathcal{CN}(0, \sigma_k^2)$ is the AWGN at user $k$ with a power of $\sigma_k^2$. The signal-to-interference-plus-noise ratio (SINR) for user $k$ to decode its own signal $s_k$ is given by
\begin{equation}\label{sec3A_eq4}
	\gamma_k(\bm{{\rm W}}, \bm{{\rm X}}) = \frac{\vert \bm{{\rm h}}_k^{{\rm H}}(\bm{{\rm X}})\bm{{\rm G}}(\bm{{\rm X}})\bm{{\rm w}}_k \vert^2}{\sum_{i=1,i\neq k}^K \vert \bm{{\rm h}}_k^{{\rm H}}(\bm{{\rm X}})\bm{{\rm G}}(\bm{{\rm X}})\bm{{\rm w}}_i \vert^2 +\sigma_k^2}.
\end{equation}
Consequently, the SE $f_{{\rm SE}}$ and EE $f_{{\rm EE}}$ in the PASS-enabled $K$-user scenario can be expressed as
\begin{equation}\label{sec3A_eq5}
	f_{{\rm SE}}(\bm{{\rm W}}, \bm{{\rm X}})=\sum_{k=1}^K {\rm log}_2(1+\gamma_k(\bm{{\rm W}}, \bm{{\rm X}})),
\end{equation}
\begin{equation}\label{sec3A_eq6}
	f_{{\rm EE}}(\bm{{\rm W}}, \bm{{\rm X}}) = \frac{\sum_{k=1}^K {\rm log}_2(1+\gamma_k(\bm{{\rm W}}, \bm{{\rm X}}))}{\sum_{k=1}^K \Vert \bm{{\rm w}}_k \Vert^2 + P_f + \chi f_{{\rm SE}}(\bm{{\rm W}}, \bm{{\rm X}})},
\end{equation}
where $\sum_{k=1}^K \Vert \bm{{\rm w}}_k \Vert^2$ is the overall transmit power, and $P_f=P_{{\rm BS}}+P_{{\rm BB}}+N_{{\rm RF}}P_{{\rm RF}}+KP_{{\rm UE}}+MN\cdot P_{{\rm PA}}$.

\subsection{Problem Formulation}
According to (\ref{sec3A_eq5}) and (\ref{sec3A_eq6}), the joint SE-EE maximization problem in $K$-user scenario can be formulated as
\begin{subequations}
	\begin{align}
		(\textbf{P2-1}): \max_{\bm{{\rm W}}, \bm{{\rm X}}} & \quad [f_{{\rm SE}}(\bm{{\rm W}}, \bm{{\rm X}})]^{\beta} \times [f_{{\rm EE}}(\bm{{\rm W}}, \bm{{\rm X}})]^{1-\beta} \label{sec3B_eq1}\\
		\mathrm{s.t.} & \quad (\ref{sec2B_eq2}), (\ref{sec2B_eq3}), \label{sec3B_eq2}\\
		& \quad \sum_{k=1}^K \Vert \bm{{\rm w}}_k \Vert^2 \leq P_{{\rm T}}, \label{sec3B_eq3} \\
		& \quad \gamma_k(\bm{{\rm W}},\bm{{\rm X}})\geq \gamma^k_{{\rm th}}, \forall k, \label{sec3B_eq4}
	\end{align}
\end{subequations}
where the constraint (\ref{sec3B_eq3}) refers to the constraint on transmit power at the BS side, and $\gamma^k_{{\rm th}}$ in (\ref{sec3B_eq4}) represents the minimum SINR requirement of user $k$. In this work, the ZF beamforming strategy is adopted. To this end, we first rewritten the signal model in (\ref{sec3A_eq3}) into the following more compact form:
\begin{equation}\label{sec3B_eq5}
	\bm{{\rm y}}\!=\!\bm{{\rm H}}^H(\bm{{\rm X}})\bm{{\rm G}}(\bm{{\rm X}})\bm{{\rm W}}\bm{{\rm s}}+\bm{{\rm n}}\!=\!(\bm{{\rm G}}^H(\bm{{\rm X}})\bm{{\rm H}}(\bm{{\rm X}}))^H\bm{{\rm W}}\bm{{\rm s}}+\bm{{\rm n}},
\end{equation}
where $\bm{{\rm H}}(\bm{{\rm X}})=[\bm{{\rm h}}_1(\bm{{\rm X}}),\bm{{\rm h}}_2(\bm{{\rm X}}),\cdots,\bm{{\rm h}}_K(\bm{{\rm X}})]\in\mathbb{C}^{MN\times K}$ is the overall channel matrix, and $\bm{{\rm n}}\in \mathbb{C}^{K\times 1}$ represents the noise vector. Let $\bm{{\rm \Psi}}(\bm{{\rm X}})\triangleq \bm{{\rm G}}^H(\bm{{\rm X}})\bm{{\rm H}}(\bm{{\rm X}})\in \mathbb{C}^{M\times K}$, given that $M\geq K$, we can obtain the following ZF beamforming matrix for arbitrary $\bm{{\rm X}}$:
\begin{equation}\label{sec3B_eq6}
	\bm{{\rm W}}=\bm{{\rm \Psi}}(\bm{{\rm X}})(\bm{{\rm \Psi}}^H(\bm{{\rm X}})\bm{{\rm \Psi}}(\bm{{\rm X}}))^{-1}\bm{{\rm P}}^{\frac{1}{2}},
\end{equation}
where $\bm{{\rm P}}={\rm diag}\{P_1, P_2, \cdots, P_K\}\in \mathbb{C}^{K\times K}$ is the diagonal power control matrix, and $P_k$ is the power coefficient for user $k$. Therefore, the transmit power can be reformulated as
\begin{align}
	&\sum_{k=1}^K\Vert \bm{{\rm w}}_k \Vert^2={\rm tr}(\bm{{\rm W}}\bm{{\rm W}}^H) \nonumber\\
	&={\rm tr}\left(\bm{{\rm \Psi}}(\bm{{\rm X}})(\bm{{\rm \Psi}}^H(\bm{{\rm X}})\bm{{\rm \Psi}}(\bm{{\rm X}}))^{-1}\bm{{\rm P}}(\bm{{\rm \Psi}}^H(\bm{{\rm X}})\bm{{\rm \Psi}}(\bm{{\rm X}}))^{-1}\!\bm{{\rm \Psi}}^H(\bm{{\rm X}})\right) \nonumber\\
	&={\rm tr}\left((\bm{{\rm \Psi}}^H(\bm{{\rm X}})\bm{{\rm \Psi}}(\bm{{\rm X}}))^{-1}\bm{{\rm \Psi}}^H(\bm{{\rm X}})\bm{{\rm \Psi}}(\bm{{\rm X}})\!(\bm{{\rm \Psi}}^H(\bm{{\rm X}})\bm{{\rm \Psi}}(\bm{{\rm X}}))^{-1}\bm{{\rm P}}\right) \nonumber \\
	&={\rm tr}\left((\bm{{\rm \Psi}}^H(\bm{{\rm X}})\bm{{\rm \Psi}}(\bm{{\rm X}}))^{-1}\bm{{\rm P}}\right). \label{sec3B_eq7}
\end{align}

Substituting (\ref{sec3B_eq7}) into (\ref{sec3A_eq4}) yields $\gamma_k=P_k/{\sigma_k^2}$. Then, problem (\textbf{P2-1}) becomes
\begin{subequations}
	\begin{align}
		(\textbf{P2-2}): \max_{\bm{{\rm P}}, \bm{{\rm X}}} & \quad \beta {\rm ln}\left(f_{{\rm SE}}(\bm{{\rm P}}, \bm{{\rm X}}) \right) \!\!+\!\! (1\!-\!\beta){\rm ln}\left(f_{{\rm EE}}(\bm{{\rm P}}, \bm{{\rm X}})\right) \label{sec3B_eq8}\\
		\mathrm{s.t.} & \quad (\ref{sec2B_eq2}), (\ref{sec2B_eq3}), \label{sec3B_eq9}\\
		& \quad {\rm tr}\left((\bm{{\rm \Psi}}^H(\bm{{\rm X}})\bm{{\rm \Psi}}(\bm{{\rm X}}))^{-1}\bm{{\rm P}}\right) \leq P_T, \label{sec3B_eq10}\\
		& \quad P_k/\sigma_k^2 \geq \gamma^k_{{\rm th}}. \label{sec3B_eq11}
	\end{align}
\end{subequations}
It is observed that the power control matrix $\bm{{\rm P}}$ is still coupled with the PA positions $\bm{{\rm X}}$ through (\ref{sec3B_eq10}). In the following, we first divide the problem (\textbf{P2-2}) into two subproblems: pinching beamforming optimization with fixed power control matrix, and power optimization with fixed PA positions. Thus, an alternating iterative algorithm of these two sub-problems is presented to get the sub-optimal solutions for problem (\textbf{P2-2}).

\subsection{Optimization of PA Positions}
When given the power control matrix $\bm{{\rm P}}$, the sub-problem for optimizing pinching beamforming can be written as
\begin{subequations}
	\begin{align}
		(\textbf{P2-3}): \min_{\bm{{\rm X}}} & \quad {\rm tr}\left((\bm{{\rm \Psi}}^{\rm H}(\bm{{\rm X}})\bm{{\rm \Psi}}(\bm{{\rm X}}))^{-1}\bm{{\rm P}}\right) \label{sec3C_eq1}\\
		\mathrm{s.t.} & \quad (\ref{sec2B_eq2}), (\ref{sec2B_eq3}), (\ref{sec3B_eq10}). \label{sec3C_eq2}
	\end{align}
\end{subequations}
The objective function in (\textbf{P2-3}) takes the form of (\ref{sec3C_eq1}) because, for a given $\bm{{\rm P}}$, expanding (\ref{sec3B_eq8}) reveals that it depends solely on the term ${\rm tr}((\bm{{\rm \Psi}}^{\rm H}(\bm{{\rm X}})\bm{{\rm \Psi}}(\bm{{\rm X}}))^{-1}\bm{{\rm P}})$. Therefore, minimizing ${\rm tr}((\bm{{\rm \Psi}}^{\rm H}(\bm{{\rm X}})\bm{{\rm \Psi}}(\bm{{\rm X}}))^{-1}\bm{{\rm P}})$ is equivalent to maximizing the original objective function (\ref{sec3B_eq8}). For the inequality constraint (\ref{sec3B_eq10}), we can integrate it into the objective function using the penalty method. Hence, problem (\textbf{P2-3}) can be reformulated as
\begin{subequations}
	\begin{align}
		(\textbf{P2-4}): \min_{\bm{{\rm X}}} & \quad {\rm tr}\left((\bm{{\rm \Psi}}^{\rm H}(\bm{{\rm X}})\bm{{\rm \Psi}}(\bm{{\rm X}}))^{-1}\bm{{\rm P}}\right)\nonumber\\
		&\quad +\tau\left(\left[{\rm tr}\left((\bm{{\rm \Psi}}^{\rm H}(\bm{{\rm X}})\bm{{\rm \Psi}}(\bm{{\rm X}}))^{-1}\bm{{\rm P}}\right)-P_T\right]^+\right)^2 \label{sec3C_eq3}\\
		\mathrm{s.t.} & \quad \quad (\ref{sec2B_eq2}), (\ref{sec2B_eq3}), \label{sec3C_eq4}
	\end{align}
\end{subequations}
where $[x]^+$ is defined as ${\rm max}(0,x)$, and $\tau$ is the penalty coefficient corresponding to the constraint (\ref{sec3B_eq10}). The problem can be solved via an element-wise one-dimensional search where in one iteration, the $x$-axis coordinate of each PA $x^p_{m,n}$ is optimized sequentially over a certain range, by keeping all other PAs fixed. However, we notice that when optimizing $x^p_{m,n}$, the matrix inversion $(\bm{{\rm \Psi}}^{\rm H}(\bm{{\rm X}})\bm{{\rm \Psi}}(\bm{{\rm X}}))^{-1}$ must be computed at each candidate solution, which results in significant computational complexity. To address this, we first present a matrix decomposition method. Recall that $\bm{{\rm \Psi}}^{\rm H}(\bm{{\rm X}})\in\mathbb{C}^{K\times M}$ can be expressed as
\begin{align}
	&\bm{{\rm \Psi}}^{\rm H}(\bm{{\rm X}})=\nonumber \\
	&\begin{bmatrix}
		\bm{{\rm h}}^{\rm H}_{1,1}(\bm{{\rm x}}_1)\bm{{\rm g}}_1(\bm{{\rm x}}_1) & \!\cdots\! & \bm{{\rm h}}^{\rm H}_{M,1}(\bm{{\rm x}}_M)\bm{{\rm g}}_M(\bm{{\rm x}}_M) \\
		\bm{{\rm h}}^{\rm H}_{1,2}(\bm{{\rm x}}_1)\bm{{\rm g}}_1(\bm{{\rm x}}_1) & \!\cdots\! & \bm{{\rm h}}^{\rm H}_{M,2}(\bm{{\rm x}}_M)\bm{{\rm g}}_M(\bm{{\rm x}}_M) \\
		\vdots & \ddots & \vdots \\
		\bm{{\rm h}}^{\rm H}_{1,K}(\bm{{\rm x}}_1)\bm{{\rm g}}_1(\bm{{\rm x}}_1) & \!\cdots\! & \bm{{\rm h}}^{\rm H}_{M,K}(\bm{{\rm x}}_M)\bm{{\rm g}}_M(\bm{{\rm x}}_M) \label{sec3C_eq5}
	\end{bmatrix}\!\!,
\end{align}
where the $k$-th element in the $m$-th column of $\bm{{\rm \Psi}}^{\rm H}(\bm{{\rm X}})$ represents the aggregate channel coefficient from all PAs on the $m$-th waveguide to the user $k$. Denote $\bm{{\rm a}}_m\in \mathbb{C}^{K\times 1}$ as the $m$-th column vector of $\bm{{\rm \Psi}}^{\rm H}(\bm{{\rm X}})$. Then, $\bm{{\rm \Psi}}^{\rm H}(\bm{{\rm X}})$ can be rewritten as $\bm{{\rm \Psi}}^{\rm H}(\bm{{\rm X}})=[\bm{{\rm a}}_1,\bm{{\rm a}}_2,\cdots,\bm{{\rm a}}_M]$, leading to the following transformation: 
\begin{equation}\label{sec3C_eq6}
	\bm{{\rm \Psi}}^{\rm H}(\bm{{\rm X}})\bm{{\rm \Psi}}(\bm{{\rm X}})=\sum_{m=1}^M \bm{{\rm a}}_m\bm{{\rm a}}_m^{\rm H},
\end{equation}
where $\bm{{\rm a}}_m\bm{{\rm a}}_m^{\rm H}\in \mathbb{C}^{K\times K}$. For optimizing the PA positions on the $m$-th waveguide, (\ref{sec3C_eq6}) can be written as
\begin{equation}\label{sec3C_eq7}
	\bm{{\rm \Psi}}^{\rm H}(\bm{{\rm X}})\bm{{\rm \Psi}}(\bm{{\rm X}})=\bm{{\rm a}}_m\bm{{\rm a}}_m^{\rm H}+\sum_{m'=1,m'\neq m}^M\bm{{\rm a}}_{m'}\bm{{\rm a}}_{m'}^{\rm H}.
\end{equation}
Let $\bm{{\rm B}}_m\triangleq \sum_{m'=1,m'\neq m}^M\bm{{\rm a}}_{m'}\bm{{\rm a}}_{m'}^{\rm H}$. Given that $M\geq K$, the matrix $\bm{{\rm B}}_m$ must be full-rank. Therefore, by using Sherman-Morrison formula, we have
\begin{align}\label{sec3C_eq8}
	&\left(\bm{{\rm \Psi}}^{\rm H}(\bm{{\rm X}})\bm{{\rm \Psi}}(\bm{{\rm X}})\right)^{-1} = (\bm{{\rm a}}_m\bm{{\rm a}}_m^{\rm H}+\bm{{\rm B}}_m)^{-1} \nonumber \\
	&=(\bm{{\rm B}}_m)^{-1}-\frac{(\bm{{\rm B}}_m)^{-1}\bm{{\rm a}}_m\bm{{\rm a}}_m^{\rm H}(\bm{{\rm B}}_m)^{-1}}{1+\bm{{\rm a}}_m^{\rm H}(\bm{{\rm B}}_m)^{-1}\bm{{\rm a}}_m}.
\end{align}
Consequently, by substituting (\ref{sec3C_eq8}) into (\ref{sec3C_eq1}), the objective function of problem (\textbf{P2-3}) can be expressed as
\begin{align}
	&{\rm tr}((\bm{{\rm \Psi}}^{\rm H}(\bm{{\rm X}})\bm{{\rm \Psi}}(\bm{{\rm X}}))^{-1}\bm{{\rm P}}) \nonumber \\
	&={\rm tr}((\bm{{\rm B}}_m)^{-1}\bm{{\rm P}})-{\rm tr}\left(\frac{(\bm{{\rm B}}_m)^{-1}\bm{{\rm a}}_m\bm{{\rm a}}_m^{\rm H}(\bm{{\rm B}}_m)^{-1}\bm{{\rm P}}}{1+\bm{{\rm a}}_m^{\rm H}(\bm{{\rm B}}_m)^{-1}\bm{{\rm a}}_m}\right)\nonumber \\
	&={\rm tr}((\bm{{\rm B}}_m)^{-1}\bm{{\rm P}})-\frac{{\rm tr}\left(\bm{{\rm a}}_m^{\rm H}(\bm{{\rm B}}_m)^{-1}\bm{{\rm P}}(\bm{{\rm B}}_m)^{-1}\bm{{\rm a}}_m\right)}{1+\bm{{\rm a}}_m^{\rm H}(\bm{{\rm B}}_m)^{-1}\bm{{\rm a}}_m} \nonumber\\
	&={\rm tr}((\bm{{\rm B}}_m)^{-1}\bm{{\rm P}})-\frac{\bm{{\rm a}}_m^{\rm H}(\bm{{\rm B}}_m)^{-1}\bm{{\rm P}}(\bm{{\rm B}}_m)^{-1}\bm{{\rm a}}_m}{1+\bm{{\rm a}}_m^{\rm H}(\bm{{\rm B}}_m)^{-1}\bm{{\rm a}}_m}. \label{sec3C_eq9}
\end{align}

Based on (\ref{sec3C_eq9}), the optimization problem for $x_{m,n}^p$ can be formulated as
\begin{subequations}
	\begin{align}
		(\textbf{P2-5}): \min_{x_{m,n}^p} & \quad f_{{\rm obj}}(x_{m,n}^p)  \label{sec3A_eq27}\\
		\mbox{s.t.} & \quad x_{m,n}^p\in \mathcal{S}_{m,n}, \label{sec3C_eq10}
	\end{align}
\end{subequations}
where $f_{{\rm obj}}(x_{m,n}^p)$ is expressed as
\begin{align}
	&f_{{\rm obj}}(x_{m,n}^p)={\rm tr}((\bm{{\rm B}}_m)^{-1}\bm{{\rm P}})-\frac{\bm{{\rm a}}_m^{\rm H}(\bm{{\rm B}}_m)^{-1}\bm{{\rm P}}(\bm{{\rm B}}_m)^{-1}\bm{{\rm a}}_m}{1+\bm{{\rm a}}_m^{\rm H}(\bm{{\rm B}}_m)^{-1}\bm{{\rm a}}_m} \nonumber\\
	&\!+\!\tau\!\!\left(\left[{\rm tr}((\bm{{\rm B}}_m)^{-1}\bm{{\rm P}})\!\!-\!\!\frac{\bm{{\rm a}}_m^{\rm H}(\bm{{\rm B}}_m)^{-1}\bm{{\rm P}}(\bm{{\rm B}}_m)^{-1}\bm{{\rm a}}_m}{1+\bm{{\rm a}}_m^{\rm H}(\bm{{\rm B}}_m)^{-1}\bm{{\rm a}}_m}\!-\!P_T\right]^+\right)^2\!\!. \label{sec3C_eq11}
\end{align}

$\mathcal{S}_{m,n}$ in constraint (\ref{sec3C_eq10}) is defined as the feasible range of $x^p_{m,n}$, which can be expressed as
\begin{equation}\label{sec3C_eq12}
	\mathcal{S}_{m,n}=[x^p_{m,n-1}+\Delta_{{\rm min}}, x^p_{m,n+1}-\Delta_{{\rm min}}] \cap [0, D_x].
\end{equation}

The decomposition method in (\ref{sec3C_eq7}) allows the inverse of matrix to be computed just once for all candidate solutions on the $m$-th waveguide, significantly reduce the computational complexity. For problem (\textbf{P2-5}), we employ particle swarm optimization (PSO) algorithm\footnote{PSO can be easily modified to support discrete PA activation.}. The PSO operates with $L$ particles with $T$ iterations, each characterized by its ``position'' and ``velocity''. The ``position'' of each particle represents a candidate solution for (\textbf{P2-5}). For brevity, the full procedure of PSO will not be restated in this paper, instead, the pinching beamforming optimization algorithm is summarized in \textbf{Algorithm} \textbf{\ref{algorithm3}}. The computational complexity of each particle is analyzed as follows. For each waveguide, the matrix inversion $(\bm{{\rm B}}_m)^{-1}$ need to be computed, resulting in a complexity of $\mathcal{O}(K^3)$. Furthermore, for each particle, the computational complexity of obtaining $\bm{{\rm a}}_m$ is $\mathcal{O}(MNK)$, the computational complexity of computing (\ref{sec3C_eq11}) is $\mathcal{O}(K^2)$. Therefore, the overall computational complexity of \textbf{Algorithm} \textbf{\ref{algorithm3}} is given by $\mathcal{O}(MK^3+TL(MNK+K^2))$.

\subsection{Optimization of Transmit Power}
To facilitate the optimization of power control matrix $\bm{{\rm P}}$, the original problem (\textbf{P2-2}) can be transformed into the following equivalent form for a given $\bm{{\rm X}}$:
\begin{subequations}
	\begin{align}
		(\textbf{P2-6}): \max_{\bm{{\rm P}}} & \quad \beta {\rm ln}\left(f_{{\rm SE}}(\bm{{\rm P}}) \right) \!+\! (1\!-\!\beta){\rm ln}\left(f_{{\rm EE}}(\bm{{\rm P}})\right) \label{sec3D_eq1}\\
		\mathrm{s.t.} & \quad {\rm tr}\left(\bm{{\rm \Lambda}}\bm{{\rm P}}\right) \leq P_T, \label{sec3D_eq2}\\
		& \quad P_k/\sigma_k^2 \geq \gamma^k_{{\rm th}}, \label{sec3D_eq3}
	\end{align}
\end{subequations}
where $\Lambda \triangleq (\bm{{\rm \Psi}}^H(\bm{{\rm X}})\bm{{\rm \Psi}}(\bm{{\rm X}}))^{-1}$. It is evident that all constraints are convex, so we only need to tackle the non-convexity of the objective function (\ref{sec3D_eq1}). First, we introduce two slack variables $\mu_1$ and $\mu_2$ such that
\begin{equation}\label{sec3D_eq4}
	\beta {\rm ln}\left(f_{{\rm SE}}(\bm{{\rm P}}) \right)\geq \mu_1,
\end{equation}
\begin{equation}\label{sec3D_eq5}
	(1\!-\!\beta){\rm ln}\left(f_{{\rm EE}}(\bm{{\rm P}})\right)\geq \mu_2.
\end{equation}
With the help of these two slack variables, problem (\textbf{P2-6}) can be equivalently written as
\begin{subequations}
	\begin{align}
		(\textbf{P2-7}): \max_{\bm{{\rm P}},\mu_1,\mu_2} & \quad \mu_1+\mu_2 \label{sec3D_eq6}\\
		\mathrm{s.t.} & \quad (\ref{sec3D_eq2}), (\ref{sec3D_eq3}),(\ref{sec3D_eq4}), (\ref{sec3D_eq5}). \label{sec3D_eq7}
	\end{align}
\end{subequations}
It can be observed that now (\ref{sec3D_eq6}) is a linear function in terms of $\mu_1$ and $\mu_2$. For constraint (\ref{sec3D_eq4}), it can be rewritten as
\begin{equation}\label{sec3D_eq8}
	{\rm e}^{\frac{\mu_1}{\beta}}-\sum_{k=1}^K{\rm log}_2\left(1+\frac{P_k}{\sigma_k^2}\right)\leq 0.
\end{equation}
It is evident that (\ref{sec3D_eq8}) is convex when $\beta>0$.\footnote{It should be noted that when $\beta=0$, (\textbf{P2-6}) reduces to the EE maximization problem, and hence, constraint (\ref{sec3D_eq4}) ceases to exist.} For (\ref{sec3D_eq5}), we have
\begin{equation}\label{sec3D_eq9}
	\frac{\sum_{k=1}^K{\rm log}_2\left(1+\frac{P_k}{\sigma_k^2}\right)}{{\rm tr}\left(\bm{{\rm \Lambda}}\bm{{\rm P}}\right)+P_f+\chi \sum_{k=1}^K{\rm log}_2\left(1+\frac{P_k}{\sigma_k^2}\right)}\geq {\rm e}^{\frac{\mu_2}{1-\beta}}.
\end{equation}
To handle the non-convexity of (\ref{sec3D_eq9}), we introduce a new slack variable $\kappa$ such that
\begin{equation}\label{sec3D_eq10}
	\frac{\sum_{k=1}^K{\rm log}_2\left(1+\frac{P_k}{\sigma_k^2}\right)}{{\rm tr}\left(\bm{{\rm \Lambda}}\bm{{\rm P}}\right)+P_f+\chi \sum_{k=1}^K{\rm log}_2\left(1+\frac{P_k}{\sigma_k^2}\right)}\!\geq\! \frac{{\rm e}^{\frac{\mu_2}{1-\beta}}\kappa}{\kappa}.
\end{equation}
Therefore, the original constraint (\ref{sec3D_eq5}) can be split into the following two constraints:
\begin{equation}\label{sec3D_eq11}
	\sum_{k=1}^K{\rm log}_2\left(1+\frac{P_k}{\sigma_k^2}\right)\geq {\rm e}^{\frac{\mu_2}{1-\beta}}\kappa,
\end{equation}
\begin{equation}\label{sec3D_eq12}
	{\rm tr}\left(\bm{{\rm \Lambda}}\bm{{\rm P}}\right)+P_f+\chi \sum_{k=1}^K{\rm log}_2\left(1+\frac{P_k}{\sigma_k^2}\right) \leq \kappa.
\end{equation}

For (\ref{sec3D_eq11}), the variables $\mu_2$ and $\kappa$ are coupled, but we notice that $u(\mu_2,\kappa)={\rm e}^{\frac{\mu_2}{1-\beta}}\kappa$ is a twice-differentiable function when $\beta\neq 1$\footnote{When $\beta=1$, (\textbf{P2-6}) reduces to the SE maximization problem, and constraint (\ref{sec3D_eq5}) ceases to exist.}, its second-order Taylor expansion around a given point $(\mu_2^{(l)}, \kappa^{(l)})$ can be expressed as
\begin{align}
	&u(\mu_2,\kappa)\!\approx\! {\rm e}^{\frac{\mu_2^{(l)}}{1-\beta}}\kappa^{(l)}\!\!+\!\!\frac{1}{1\!-\!\beta}{\rm e}^{\frac{\mu_2^{(l)}}{1\!-\!\beta}}\kappa^{(l)}(\mu_2\!-\!\mu_2^{(l)})\!+\!{\rm e}^{\frac{\mu_2^{(l)}}{1\!-\!\beta}}(\kappa\!-\!\kappa^{(l)}) \nonumber \\
	&\!\!+\!\!\frac{1}{2}\left[\left[\mu_2\!-\!\mu_2^{(l)},\kappa\!-\!\kappa^{(l)} \right]\!\nabla^2u(\mu_2^{(l)},\kappa^{(l)})\left[\mu_2\!-\!\mu_2^{(l)}, \kappa\!-\!\kappa^{(l)}\right]^{\rm T}\right]\!,\label{sec3D_eq13}
\end{align}

\noindent where $l$ is the iteration index of successive convex approximation (SCA). However, if we simply substitute eq. (\ref{sec3D_eq13}) into constraint (\ref{sec3D_eq11}), the obtained solution may not satisfy the constraints of the original problem. According to the Descent Lemma \cite{bertsekas1999nonlinear, sun2017majorization}, by upper-bounding the Hessian matrix via $\nabla^2u(\mu_2^{(l)},\kappa^{(l)}) \preceq \delta\bm{{\rm I}}$, where $\delta\in \mathbb{R}_+$, we can get the convex upper bound of $u(\mu_2,\kappa)$ as
\begin{align}
	u(\mu_2,\kappa)&\leq {\rm e}^{\frac{\mu_2^{(l)}}{1-\beta}}\kappa^{(l)}\!\!+\!\!\frac{1}{1\!-\!\beta}{\rm e}^{\frac{\mu_2^{(l)}}{1\!-\!\beta}}\kappa^{(l)}(\mu_2\!-\!\mu_2^{(l)})\!+\!{\rm e}^{\frac{\mu_2^{(l)}}{1\!-\!\beta}}(\kappa\!-\!\kappa^{(l)}) \nonumber \\
	&+\frac{\delta}{2}\left[(\mu_2-\mu_2^
	{(l)})^2+(\kappa-\kappa^{(l)})^2\right]=U^{(l)}(\mu_2,\kappa).\label{sec3D_eq14}
\end{align}
\noindent In (\ref{sec3D_eq14}), we set $\delta$ as Frobenius norm of the Hessian matrix $\nabla^2u(\mu_2^{(l)},\kappa^{(l)})$\cite{liu2025joint}. Consequently, the original constraint (\ref{sec3D_eq11}) now becomes $U^{(l)}(\mu_2, \kappa)-\sum_{k=1}^K{\rm log}_2\left(1+P_k/\sigma_k^2\right)\leq 0$, which is convex.

For constraint (\ref{sec3D_eq12}), the function ${\rm log}_2(1+P_k/\sigma_k^2)$ is concave with respect to $P_k$. Since any concave function is globally upper-bounded by its first-order Taylor expansion at any point, we can derive an upper bound for ${\rm log}_2(1+P_k/\sigma_k^2)$ at the given local point $P_k^{(l)}$ in the $l$-th iteration as
\begin{equation}\label{sec3D_eq15}
	{\rm log}_2(1\!+\!\frac{P_k}{\sigma_k^2}) \!\leq\! {\rm log}_2(1\!+\!\frac{P_k^{(l)}}{\sigma_k^2})\!+\!\frac{P_k\!-\!P_k^{(l)}}{({\rm ln}2)(\sigma_k^2\!\!+\!\!P_k^{(l)})}\!=\!g_k^{(l)}(P_k),
\end{equation}
thus, we can convert (\ref{sec3D_eq12}) into the following convex constraint:
\begin{equation}\label{sec3D_eq16}
	{\rm tr}\left(\bm{{\rm \Lambda}}\bm{{\rm P}}\right)+P_f+\chi \sum_{k=1}^Kg_k^{(l)}(P_k)-\kappa\leq 0.
\end{equation}

Based on the above approximations, the problem (\textbf{P2-7}) can be further reformulated as
\begin{subequations}
	\begin{align}
		(\textbf{P2-8}): \max_{\bm{{\rm \Xi}}} & \quad \mu_1+\mu_2 \label{sec3D_eq17}\\
		\mathrm{s.t.} & \quad (\ref{sec3D_eq2}),(\ref{sec3D_eq3}),(\ref{sec3D_eq8}), \label{sec3D_eq18}\\
		& \quad U^{(l)}(\mu_2, \kappa)\!\!-\!\!\sum_{k=1}^K{\rm log}_2\!\!\left(1\!+\!\frac{P_k}{\sigma_k^2}\right)\!\leq\!\! 0, \label{sec3D_eq19}\\
		& \quad {\rm tr}\left(\bm{{\rm \Lambda}}\bm{{\rm P}}\right)\!+\!P_f\!+\!\chi \sum_{k=1}^Kg_k^{(l)}(P_k)\!-\!\kappa\!\leq\! 0, \label{sec3D_eq20}
	\end{align}
\end{subequations}
where $\bm{{\rm \Xi}}=\{\bm{{\rm P}},\mu_1,\mu_2,\kappa\}$ consists of all the variables involved in this design. Now, problem (\textbf{P2-8}) is convex and therefore can be solved optimally by standard convex program solvers such as CVX.

\begin{algorithm}[t]
	\caption{Proposed Method for Solving (\textbf{P2-4})}
	\label{algorithm3}
	\begin{algorithmic}[1]
		\STATE \textbf{Initialize}: Set iteration index $r$=1, initial variables $\bm{{\rm X}}^{(r)}$, power control matrix $\bm{{\rm P}}$, and convergence tolerance $0\leq \epsilon_1 \ll 0$.
		\REPEAT
		\FOR{$m\in\{1,2,...,M\}$}
		\FOR{$n\in\{1,2,...,N\}$}
		\STATE Update $x_{m,n}^p$ by solving problem (\textbf{P2-5}) through PSO for given $\bm{{\rm P}}$
		\ENDFOR
		\ENDFOR
		\STATE Set $r=r+1$
		\UNTIL{The fractional decrease of the objective value of problem (\textbf{P2-4}) is below a threshold $\epsilon_1$}
	\end{algorithmic}
\end{algorithm}

\begin{algorithm}[t]
	\caption{ZF-based BCD Algorithm for Solving (\textbf{P2-2})}
	\label{algorithm4}
	\begin{algorithmic}[1]
		\STATE \textbf{Initialize}: Set iteration index $q$=1, initial variables $\bm{{\rm X}}^{(q)}$, $\mu_2^{(q)}$, $\kappa^{(q)}$, $\bm{{\rm P}}^{(q)}$, and convergence tolerance $0\leq \epsilon_2 \ll 0$.
		\REPEAT
		\STATE Set iteration index $l$=1, error tolerance $0\leq \epsilon_3\ll 1$
		\STATE Set $\mu_2^{(l)}=\mu_2^{(q)}$, $\kappa^{(l)}=\kappa^{(q)}$, and $\bm{{\rm P}}^{(l)}=\bm{{\rm P}}^{(q)}$
		\REPEAT
		\STATE Solve (\textbf{P2-8}) for given $\bm{{\rm X}}^{(q)}$,$\mu_2^{(l)}$,$\kappa^{(l)}$ and $\bm{{\rm P}}^{(l)}$
		\STATE Updating $\mu_2^{(l+1)}$,$\kappa^{(l+1)}$ and $\bm{{\rm P}}^{(l+1)}$
		\STATE Set $l=l+1$
		\UNTIL{The fractional increase of the objective value of problem (\textbf{P2-8}) is below a threshold $\epsilon_3$}
		\STATE Denote the optimal power control matrix as $\bm{{\rm P}}^{(q+1)}$
		\STATE Obtain the optimal positions of PASS $\bm{{\rm X}}^{(q+1)}$ by using \textbf{Algorithm} \textbf{\ref{algorithm3}} for given $\bm{{\rm P}}^{(q+1)}$
		\STATE Set $q=q+1$
		\UNTIL{The fractional increase of the objective value of problem (\textbf{P2-2}) is below a threshold $\epsilon_2$}
	\end{algorithmic}
\end{algorithm}

\vspace{-\baselineskip}
\subsection{Overall Algorithm and Convergence}
Building on the results from previous two subsections, we present an overall iterative algorithm for problem (\textbf{P2-2}) based on block coordinate descent (BCD) method. Specifically, the entire optimization variables in original problem (\textbf{P2-2}) are partitioned into two blocks, i.e., $\{\bm{{\rm P}},\bm{{\rm X}}\}$. Then, the power control matrix $\bm{{\rm P}}$ and the PA positions $\bm{{\rm X}}$ are alternately optimized, by solving problem (\textbf{P2-8}) and (\textbf{P2-4}) correspondingly, while keeping the other variable fixed. The obtained solution from each iteration is used as the input in the subsequent one. The ZF-based BCD algorithm for problem (\textbf{P2-2}) is summarized in \textbf{Algorithm} \textbf{\ref{algorithm4}}.

Next, we provide the analysis of the convergence of \textbf{Algorithm} \textbf{\ref{algorithm4}}. Since a new slack variable $\kappa$ is introduced when handling problem (\textbf{P2-7}), we define an intermediate optimization problem (\textbf{P2-7-1}) as
\begin{subequations}
	\begin{align}
		(\textbf{P2-7-1}): \max_{\bm{{\rm P}},\mu_1,\mu_2,\kappa} & \quad \mu_1+\mu_2 \label{sec3E_eq1}\\
		\mathrm{s.t.} & \quad (\ref{sec3D_eq2}),(\ref{sec3D_eq3}),(\ref{sec3D_eq8}), (\ref{sec3D_eq11}), (\ref{sec3D_eq12}). \label{sec3E_eq2}
	\end{align}
\end{subequations}
It is noticed that problem (\textbf{P2-6}), (\textbf{P2-7}) and (\textbf{P2-7-1}) are equivalent. For given $\bm{{\rm X}}^{(q)}$, $\bm{{\rm P}}^{(q)}$, $\mu_1^{(q)}$, $\mu_2^{(q)}$ and $\kappa^{(q)}$ in steps 3-10 of \textbf{Algorithm} \textbf{\ref{algorithm4}}, we have
\begin{align}
	&\mathcal{F}_{(\bm{{\rm P2-2}})}\left(\bm{{\rm X}}^{(q)},\bm{{\rm P}}^{(q)}\right)\stackrel{(a)}{=}\mathcal{F}_{(\bm{{\rm P2-6}})}\left(\bm{{\rm X}}^{(q)},\bm{{\rm P}}^{(q)}\right)\nonumber \\
	&=\mathcal{F}_{(\bm{{\rm P2-7-1}})}\left(\bm{{\rm X}}^{(q)},\bm{{\rm P}}^{(q)},\mu_1^{(q)}, \mu_2^{(q)},\kappa^{(q)}\right) \nonumber \\
	&\stackrel{(b)}{=}\mathcal{F}_{(\bm{{\rm P2-8}})}\left(\bm{{\rm X}}^{(q)},\bm{{\rm P}}^{(q)},\mu_1^{(q)}, \mu_2^{(q)},\kappa^{(q)}\right) \nonumber \\
	&\stackrel{(c)}{\leq}\mathcal{F}_{(\bm{{\rm P2-8}})}\left(\bm{{\rm X}}^{(q)},\bm{{\rm P}}^{(q+1)},\mu_1^{(q+1)}, \mu_2^{(q+1)},\kappa^{(q+1)}\right) \nonumber \\
	&\stackrel{(d)}{\leq}\mathcal{F}_{(\bm{{\rm P2-7-1}})}\left(\bm{{\rm X}}^{(q)},\bm{{\rm P}}^{(q+1)},\mu_1^{(q+1)}, \mu_2^{(q+1)},\kappa^{(q+1)}\right) \nonumber \\
	&=\mathcal{F}_{(\bm{{\rm P2-6}})}\left(\bm{{\rm X}}^{(q)},\bm{{\rm P}}^{(q+1)}\right)\stackrel{(e)}{=}\mathcal{F}_{(\bm{{\rm P2-2}})}\left(\bm{{\rm X}}^{(q)},\bm{{\rm P}}^{(q+1)}\right) \nonumber\\
	&\stackrel{(f)}{\leq}\mathcal{F}_{(\bm{{\rm P2-2}})}\left(\bm{{\rm X}}^{(q+1)},\bm{{\rm P}}^{(q+1)}\right),\label{sec3E_eq3}
\end{align}
where $\mathcal{F}_{(\cdot)}$ represents the objective value of different problem. (a) and (e) hold since (\textbf{P2-2}) and (\textbf{P2-6}) have the same objective function value at any feasible $\bm{{\rm P}}$ with given PA positions $\bm{{\rm X}}$, (b) holds the Taylor expansions in (\ref{sec3D_eq14}) and  (\ref{sec3D_eq15}) are tight at the given local points, (c) holds since in steps 3-10 of \textbf{Algorithm} \textbf{\ref{algorithm4}}, problem (\textbf{P2-8}) is solved optimally with solution $\{\bm{{\rm P}}^{(q+1)},\mu_1^{(q+1)}, \mu_2^{(q+1)},\kappa^{(q+1)}\}$ under the given $\bm{{\rm X}}^{(q)}$, (d) holds since the objective value of problem (\textbf{P2-8}) is the lower bound of that of problem (\textbf{P2-7-1}), and (f) holds since the monotonic convergence is guaranteed in PSO method when the final solution satisfy the constraint (\ref{sec3B_eq10}). On the other hand, the objective value of problem (\textbf{P2-2}) is upper bounded by a finite value. Therefore, the proposed \textbf{Algorithm} \textbf{\ref{algorithm4}} is guaranteed to converge.

\section{Numerical Results}\label{sec:numerical results}

In this section, numerical results are provided to illustrate the effectiveness of our proposed SE-EE tradeoff design of a PASS-assisted downlink system. Without loss of generality, we consider the waveguides are deployed at a height of $h$=3 m. Some of the other simulation parameters are set as follows: the noise power is $-90$ dBm\cite{xu2025rate}, the minimum SNR of all users is set to 6 dB, $f_c$ is 28 GHz, $\Delta_{{\rm min}}=\lambda/2$, $n_{{\rm neff}}=1.4$, and $\chi$=0.1. For PSO, we employ a swarm of 30 particles over 300 iterations, with inertia weight, penalty coefficient $\tau$, cognitive, and social parameters set to 0.7298, $10^4$, 1.4962 and 1.4962, respectively. Unless stated otherwise, the detailed simulation setup is given in Table \ref{table1}.

\begin{table}[t]
	\caption{Simulation Parameters}
	\label{table1}
	\centering
	\begin{tabular}{|c|c|}
		\hline
		Height of the waveguide $h$ & 3 m\cite{xu2025rate}\\
		Carrier frequency $f_c$ & 28 GHz\cite{xu2025rate}\\
		Minimum PA spacing $\Delta_{{\rm min}}$ & $\lambda/2$\cite{xu2025rate}\\
		Waveguide effective refractive index $n_{{\rm eff}}$ & 1.4\cite{xu2025rate}\\
		In-waveguide attenuation $a$ & 0.0092 m$^{-1}$\cite{xu2025pinching}\\
		Noise power & -90 dBm\cite{xu2025rate}\\
		$P_{{\rm BS}}$ & 3 W\cite{wang2023simultaneously}\\
		$P_{{\rm BB}}$ & 0.3 W\cite{wang2023simultaneously}\\
		$P_{{\rm RF}}$ & 0.2 W\cite{wang2023simultaneously}\\
		$P_{{\rm UE}}$ & 0.1 W\cite{wang2023simultaneously}\\
		$P_{{\rm PA}}^{{\rm act}}$ & 5 dBm\cite{gan2025dual} \\
		$P_{{\rm PA}}^{{\rm mot}}$ & 20 dBm\cite{gan2025dual}\\
		$P_{{\rm PA}}^{{\rm pie}}$ & 8 dBm\cite{gan2025dual}\\
		$\chi$ & 0.1 W/(bit/s/Hz)\cite{zhou2025sum}\\
		Number of particles & 30\cite{kim2017topology}\\
		Number of PSO iterations & 300\cite{kim2017topology} \\
		Inertia weight & 0.7298\cite{kim2017topology}\\
		Penalty coefficient $\tau$ & $10^4$\\
		Cognitive parameter & 1.4962\cite{kim2017topology}\\
		Social parameter & 1.4962\cite{kim2017topology}\\
		SNR threshold $\gamma_{{\rm th}}$ & 6 dB\\
		\hline
	\end{tabular}
\end{table}

\begin{figure}[t]
	\centering
	\includegraphics[width=3.5in]{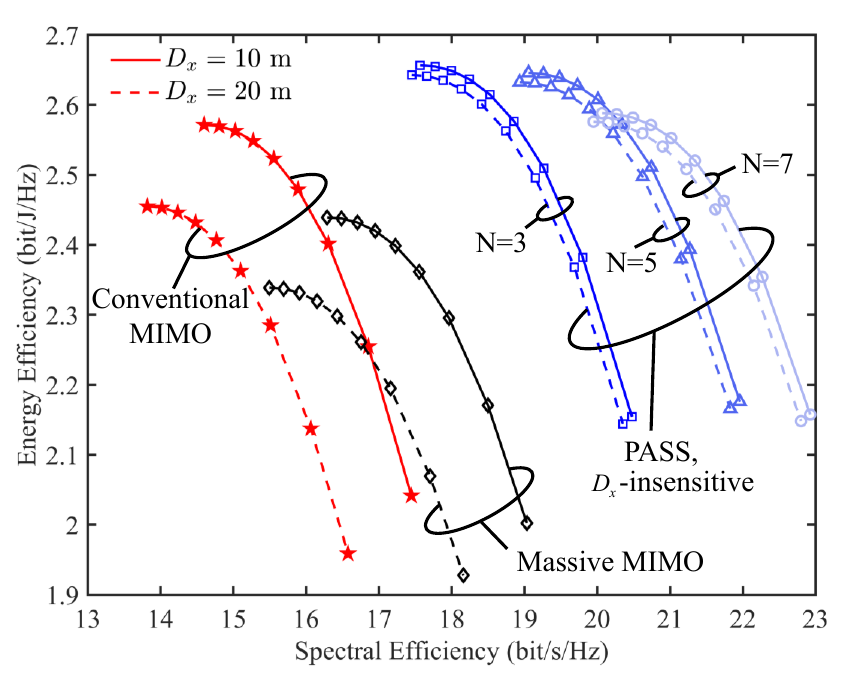}
	\caption{The SE-EE trade-off in single-user case under different $D_x$ and $N$.}
	\label{figure_6}
\end{figure}

\subsection{Single-User System}
\begin{figure}[t]
	\centering
	\includegraphics[width=3.5in]{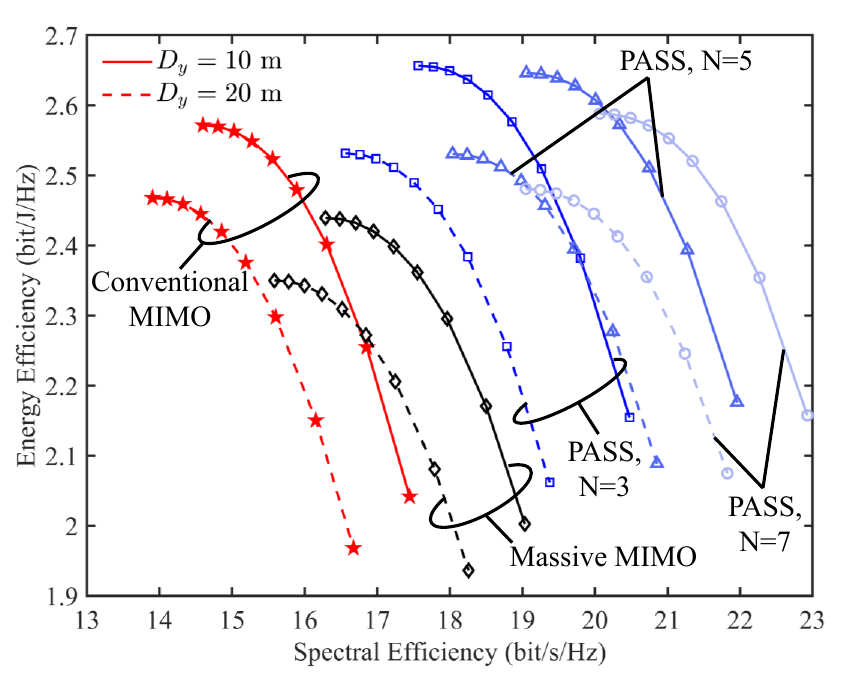}
	\caption{The SE-EE trade-off in single-user case under different $D_y$ and $N$.}
	\label{figure_7}
\end{figure}

\begin{figure}[t]
	\centering
	\includegraphics[width=3.5in]{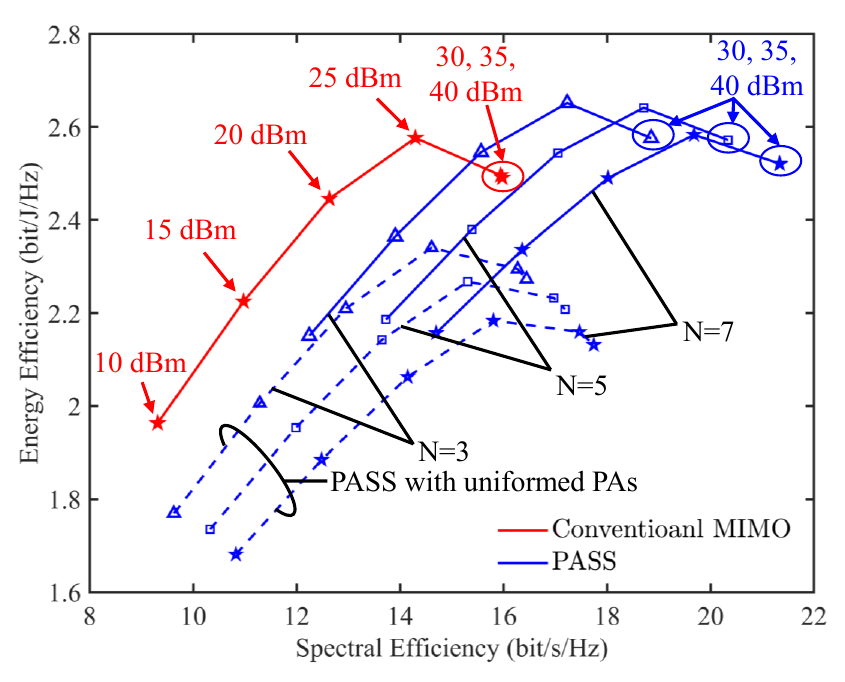}
	\caption{SE and EE performance in single-user case under different $P_T$, with $\beta$=0.5.}
	\label{figure_8}
\end{figure}

Fig. \ref{figure_6} and \ref{figure_7} illustrate the SE-EE tradeoff in a single-user scenario with different map shapes.\footnote{For fairness, the MIMO BS is always deployed at the center of the service region. The massive MIMO system is equipped with $MN$ antennas, each with its own RF chain\cite{shan2025exploiting}, in Fig. \ref{figure_6} and \ref{figure_7}, $M, N$ are set to 2 and 2 for massive MIMO to avoid visual clutter.} Each point on the curve corresponds to the SE and EE pair of the optimal solutions obtained at a specific $\beta$, which takes values from 0 to 1 in steps of 0.1. First, it is observed that PASS ($M=2$) can significantly improves the SE of the system compared to conventional MIMO and massive MIMO systems, and this improvement becomes more pronounced as $N$ increase. A notable observation is that under the same settings, PASS at $\beta$=0 can even achieve a higher SE than conventional MIMO at $\beta$=1. Second, we notice that increasing the number of PAs to improve SE comes at the expense of reduced EE, since more PAs means higher power consumption on PA activation and motor modules. However, when $M=2, N=\{3, 5\}$, which are the cases demonstrated in Fig. \ref{figure_6} and \ref{figure_7}, PASS consistently outperforms conventional and massive MIMO systems in terms of the joint SE-EE performance under different $\beta$. For massive MIMO system equipped with $MN$ antennas, it exhibits a significant lower EE due to the high energy costs from its large number of RF chains. Another interesting observation is in single-user scenario, PASS demonstrates robustness to variations in the length of the service area in waveguide direction, i.e., $D_x$ in this paper. While the performance of MIMO systems degrade significantly with a larger $D_x$, PASS exhibits insensitivity to this parameter. This is mainly because when $D_y$ is fixed and $D_x, D_y$ meet certain conditions (see case 2.2 and 2.3 in Section \ref{sec:2:pa_placement}), we can flexibly adjust the PAs' position to keep them as close as possible to the user, regardless of its x-coordinate\footnote{It needs to be emphasized that if the coarse position of PAs is set according to (\ref{sec2C_eq12}), then PASS will still be sensitive to $D_x$.}. Nevertheless, both PASS and MIMO exhibit a similar level of performance degradation as $D_y$ increases. This is potentially because an increased $D_y$ leads to larger path loss, resulting in the degradation of SE and EE. Additionally, in Fig. \ref{figure_8}, we compare the SE-EE performance under $\beta$=0.5 to highlight the necessity of jointly optimizing the transmit power and the pinching beamforming. The blue dotted line represents a baseline of uniformly placed PAs under optimized transmit power. Results show that joint optimization yields better joint SE-EE performance under the same transmit power constraint $P_T$ at the BS. Moreover, we notice that even with uniform PA placement, the SE performance of the system are still better than that of the conventional MIMO, especially with larger $N$, which indicates that PASS is beneficial for boosting SE the system.

\begin{figure*}[t]
	\centering
	\includegraphics[width=6.5in]{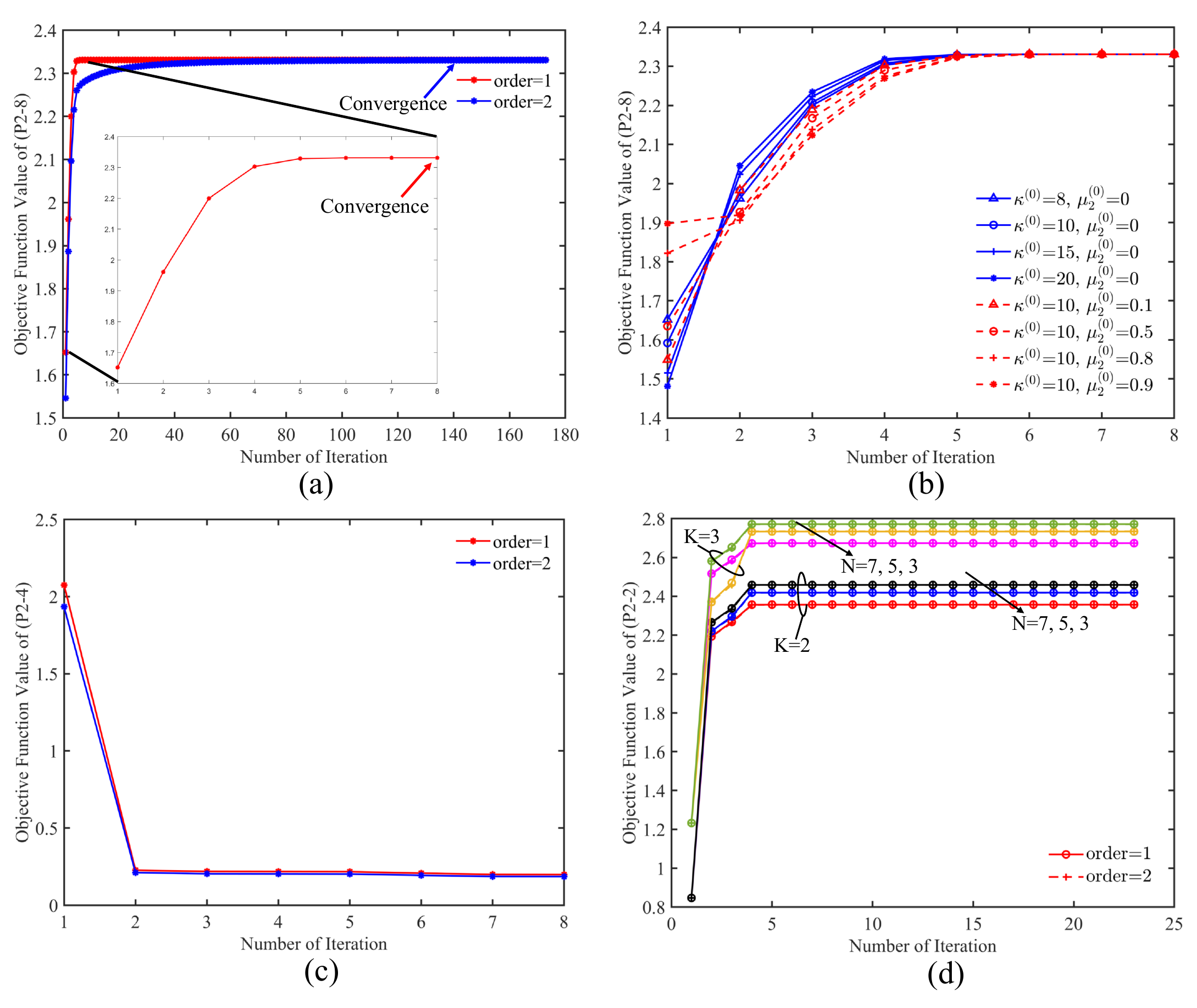}
	\caption{Convergence of (a) the objective function in (\textbf{P2-8}), (b) numerical stability of the First order Taylor expansion scheme under different initial points, (c) the objective function in (\textbf{P2-4}), and (d) the \textbf{Algorithm} \textbf{\ref{algorithm4}}.}
	\label{figure_9}
\end{figure*}

\begin{figure}[t]
	\centering
	\includegraphics[width=3.5in]{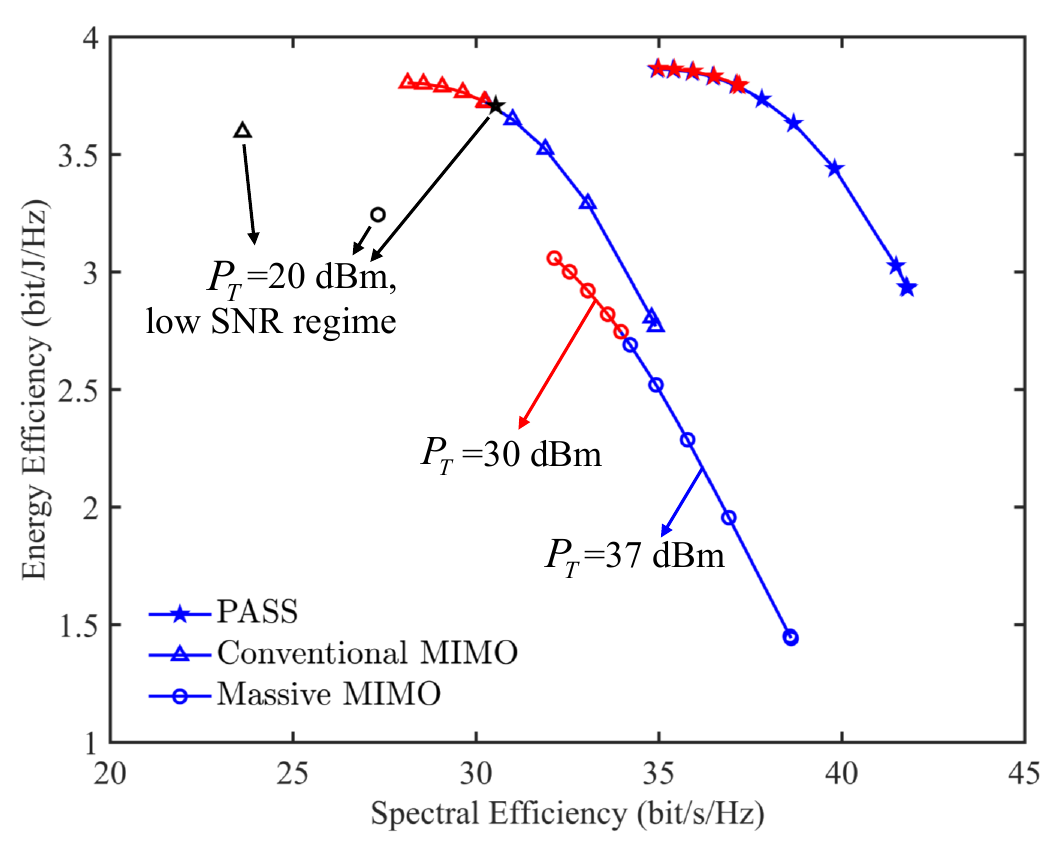}
	\caption{The SE-EE trade-off under different $P_T$.}
	\label{figure_10}
\end{figure}

\begin{figure}[t]
	\centering
	\includegraphics[width=3.5in]{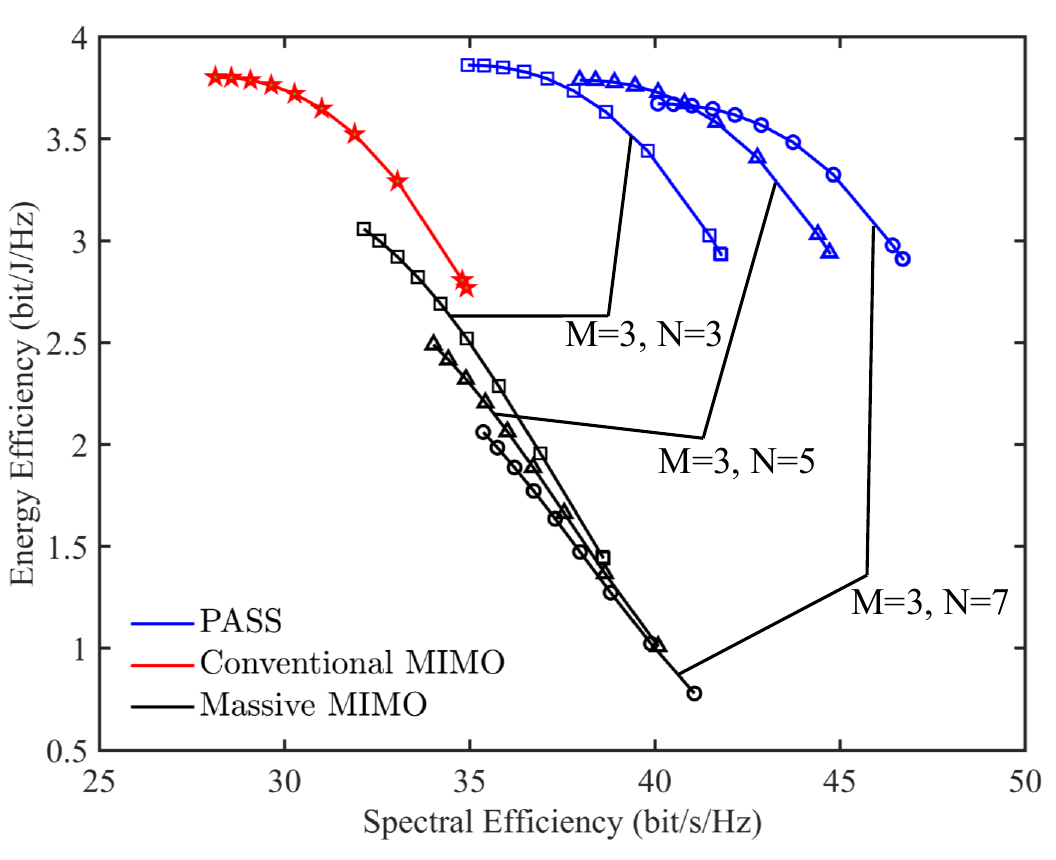}
	\caption{The SE-EE trade-off under different $P_T$.}
	\label{figure_11}
\end{figure}

\begin{figure}[t]
	\centering
	\includegraphics[width=3.5in]{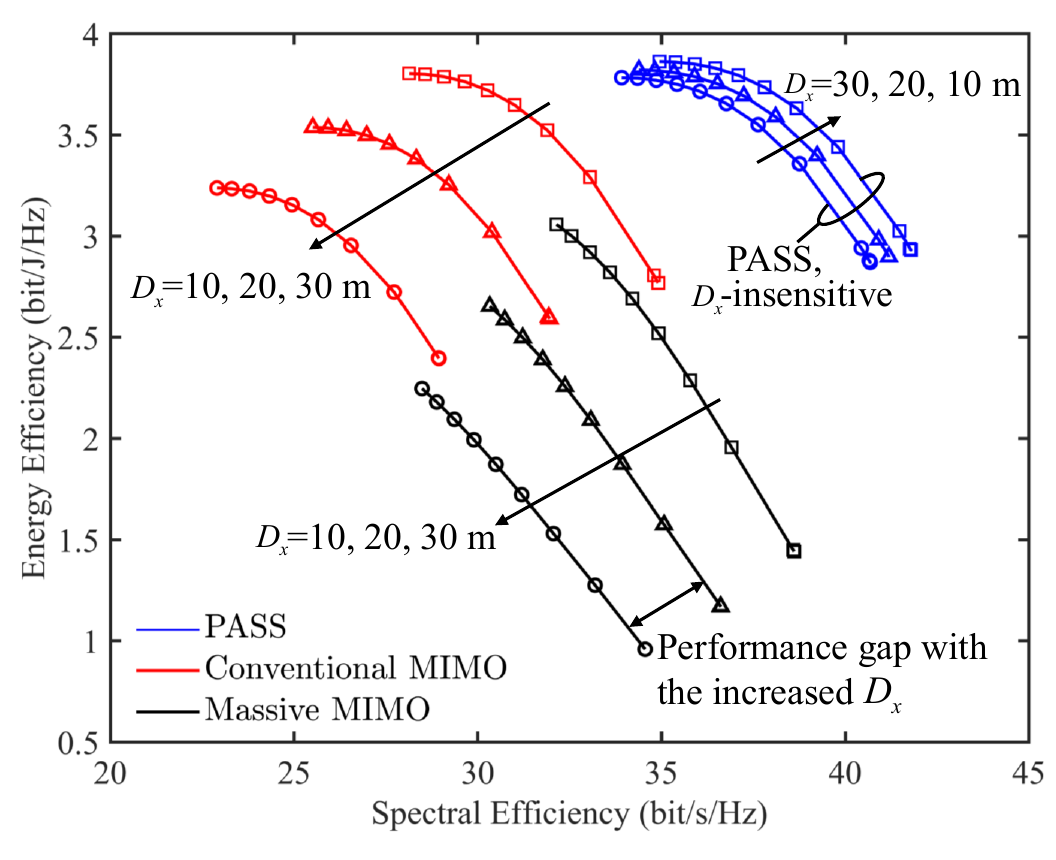}
	\caption{The SE-EE trade-off under different $D_x$.}
	\label{figure_12}
\end{figure}

\begin{figure}[t]
	\centering
	\includegraphics[width=3.5in]{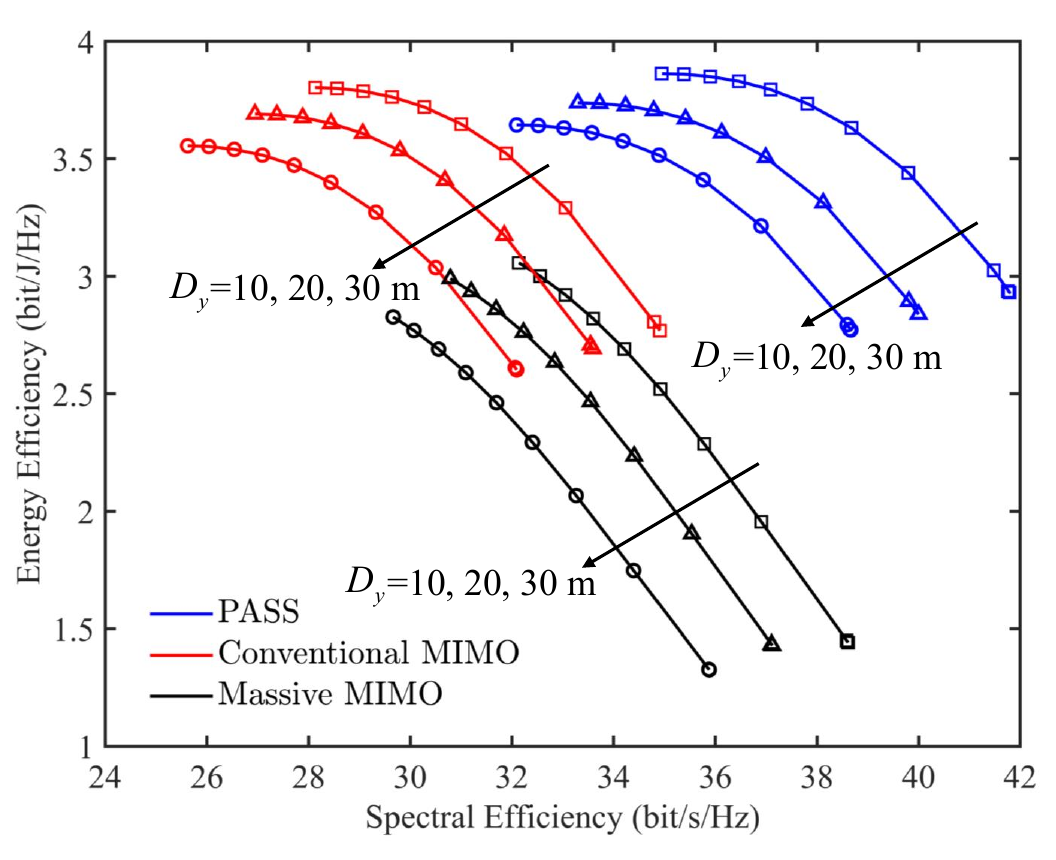}
	\caption{The SE-EE trade-off under different $D_y$.}
	\label{figure_13}
\end{figure}

\subsection{Multi-User System}
\begin{figure}[t]
	\centering
	\includegraphics[width=3.5in]{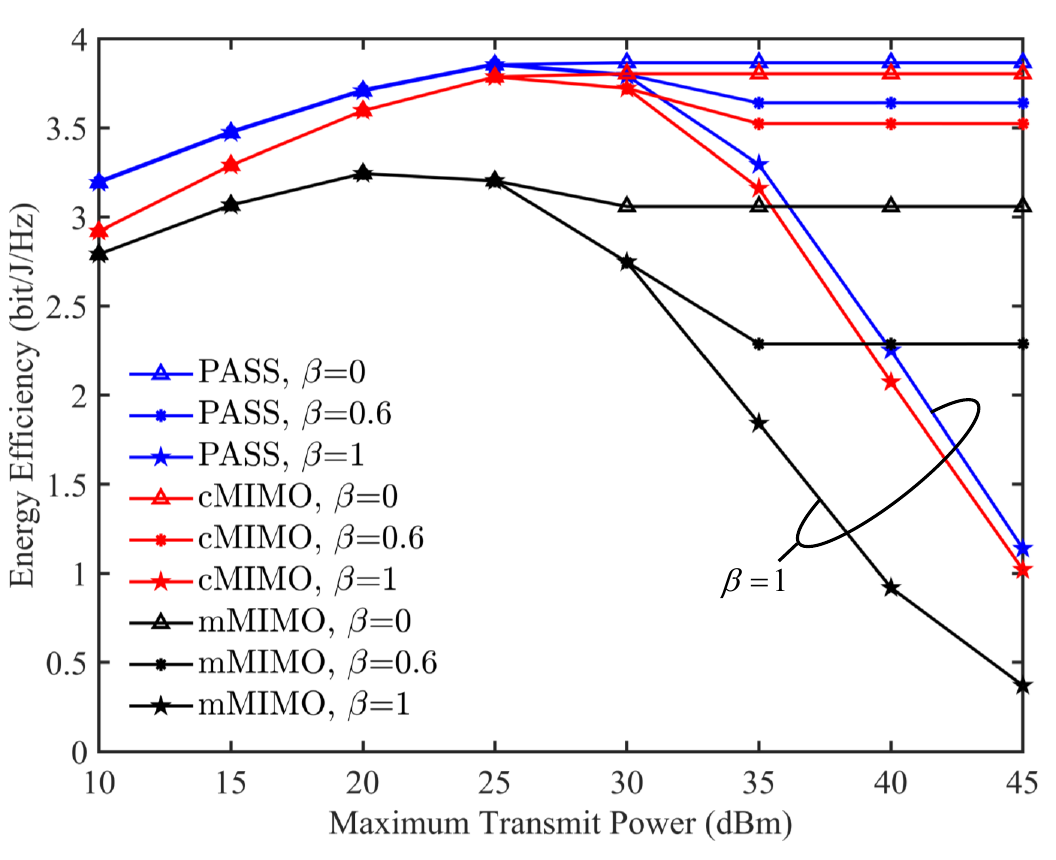}
	\caption{Achieved EE versus $P_T$ for various $\beta$ (Conventional and massive MIMO are abbreviated as cMIMO and mMIMO, respectively).}
	\label{figure_14}
\end{figure}

\begin{figure}[t]
	\centering
	\includegraphics[width=3.5in]{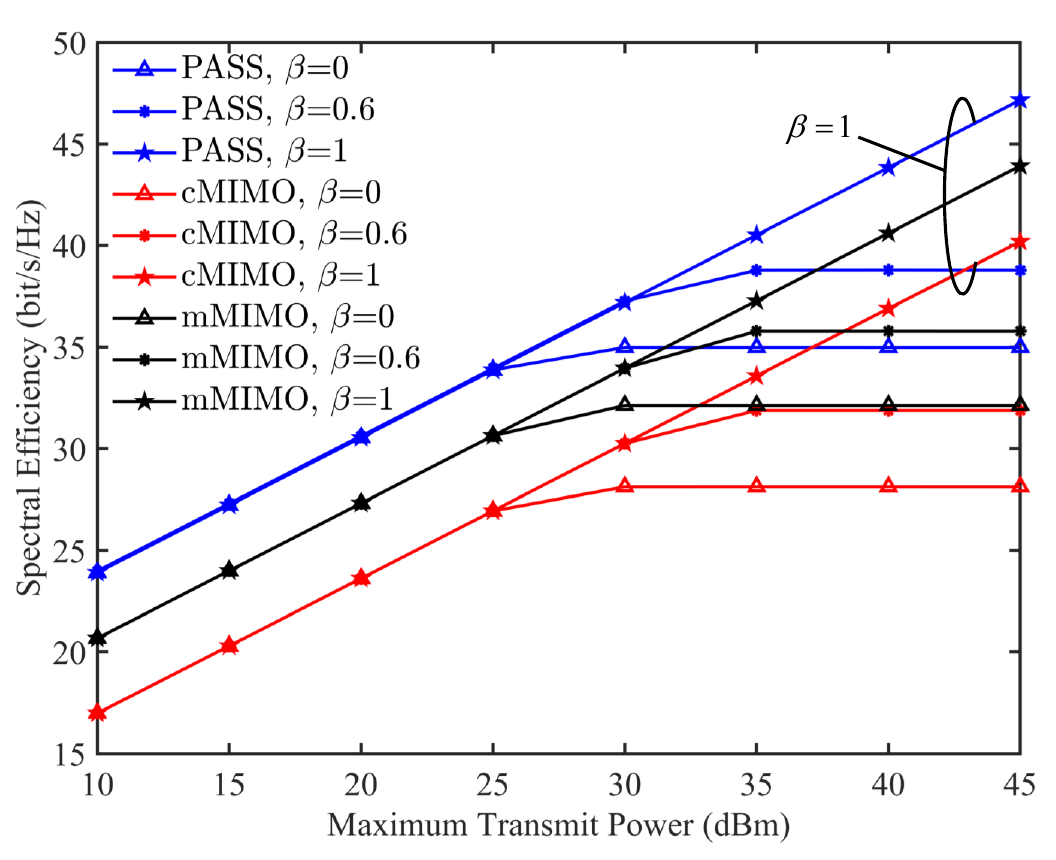}
	\caption{Achieved SE versus $P_T$ for various $\beta$ (Conventional and massive MIMO are abbreviated as cMIMO and mMIMO, respectively).}
	\label{figure_15}
\end{figure}

Fig. \ref{figure_9} illustrates the convergence behavior of the proposed algorithms. Specifically, the convergence behavior of the objective function in (\textbf{P2-8}) is shown in Fig. \ref{figure_9}(a). The red and blue curves represent different expansion schemes for $u(\mu_2,\kappa)$ in (\ref{sec3D_eq14}), corresponding to the first-order and second-order Taylor expansions, respectively. It can be observed that the first-order Taylor expansion scheme achieves significantly faster convergence speed on (\textbf{P2-8}) compared to the using second-order Taylor expansion, while both approaches converge to the same optimal value, thus, the first-order Taylor expansion scheme is adopted in the subsequent simulations, and its numerical stability under different expansion points is shown in Fig. \ref{figure_9}(b). Fig. \ref{figure_9}(c) shows the convergence behavior of the objective function in (\textbf{P2-4}), and in Fig. \ref{figure_9}(d), the convergence of \textbf{Algorithm} \textbf{\ref{algorithm4}} for different $N$ and $K$ is demonstrated. The threshold $\epsilon_2$ to terminate the algorithm has been set to $10^{-6}$. As shown in Fig. \ref{figure_9}(d), the overall algorithm converges with a small number of iterations under both first- and second-order Taylor expansions. 

Fig. \ref{figure_10} shows the SE-EE tradeoff in PASS, conventional MIMO and massive MIMO systems under different $P_T$ while keeping $D_x$=$D_y$=10 m, $M$=$N$=3, and $K$=2. When $P_T$=37 dBm, a tradeoff between SE and EE exists in three systems. However, this trade-off vanishes when $P_T$ is reduced to 20 dBm, where both SE and EE remain unchanged in the range of $\beta$ from 0 to 1. This is because the Pareto optimal set for problem (\textbf{P2-6}) degenerates to a single point $P_T$. Fig. \ref{figure_11} illustrates the SE-EE tradeoff under different number of PAs with $P_T$=37 dBm, $D_x$=$D_y$=10 m, $M$=3, and $K$=2. It can be observed that as $N$ increases, the gap in SE between PASS and the other two baselines widens gradually. Meanwhile, the EE of both PASS and massive MIMO decreases with the grown of $N$. For PASS, this is due to the additional power consumption introduced by extra PA activation and movement. For massive MIMO, the reduction in EE results from the higher power consumption caused by the increased number of RF chains. Overall, when $M$=$N$=3, PASS can achieve better joint SE-EE performance compared with conventional and massive MIMO systems. Fig. \ref{figure_12} illustrates the effect of $D_x$ on the SE-EE tradeoff with $P_T$=37 dBm, $D_y$=10 m, $M$=$N$=3, and $K$=2. It can be seen that expanding the service area significantly degrades the SE and EE performance of MIMO systems. In contrast, PASS is considerably less affected by an increase in $D_x$, due to its capacity to flexibly reposition the PAs to mitigate the large-scale path loss. This indicates that PASS is especially useful in some specific scenarios like tunnels, lone corridors, etc. Furthermore, Fig. \ref{figure_13} shows the SE-EE tradeoff under different $D_y$ while keeping $P_T$=37 dBm, $D_x$=10 m, $M$=$N$=3, and $K$=2. We can observe that when $D_y$ is increased, the performance of all three systems degrades to a similar extent. This is consistent with our observation in Fig. \ref{figure_7}, which is attributed to the higher path loss introduced by an increased $D_y$, leading to the decline in both SE and EE.

Fig. \ref{figure_14} illustrates the achieved EE of the system versus $P_T$ for various weighting coefficients $\beta$. Specifically, when $\beta$=0, the original problem becomes a EE-maximization design for both PASS and MIMO system. Take PASS as an example, its EE increases initially and then remains constant. On the other hand, when $\beta$=1, the original problem turns into a SE-maximization design. In contrast to the EE-maximization counterpart, the EE of both PASS and MIMO systems starts to decline at higher available power levels since SE-maximization design aims to maximize the sum rate at the cost of EE degradation, but it is worth noting that for $\forall \beta\in[0,1]$, PASS always achieves better EE performance compared with the other two baselines, which reflects that PASS is an energy-efficient architecture. Fig. \ref{figure_15} shows the SE comparison versus $P_T$ for various $\beta$. When $\beta$=1, the original SE-EE tradeoff design becomes conventional SE-maximization, in this case, the SE of both PASS and MIMO are increasing linearly with $P_T$ under ZF beamforming strategy. When $\beta$=0, the original problem turns out to be a EE-maximization design in which SE increases with $P_T$ until it reaches the ``\textit{green power}'' as described in \cite{Obiedollah2025energy}, and then SE and EE saturate. Similarly, it is also observed that under the same setting, PASS always achieves better SE performance compared with MIMO under different $P_T$ and $\beta$.

\section{Conclusion}\label{sec:conclusion}
A joint transmit and pinching beamforming design to address the SE-EE tradeoff in both PASS-enabled single- and multi-user scenarios were investigated. In the single-user scenario, a two-stage joint beamforming design is proposed: In the first stage, a general PA placement framework is proposed. In the second stage, the closed-form solution for the optimal transmit power was derived with the given optimized PA positions under the MRT strategy. In multi-user scenario, a ZF-assisted AO algorithm was proposed to tackle the non-convex joint transmit and pinching beamforming problem. The power allocation subproblem was addressed by deriving the convex upper bounds of the constraints, and the pinching beamforming was optimized through an iterative element-wise sequential PSO method. The numerical results suggested that the number of waveguides and PAs should be carefully configured to achieve better joint SE-EE performance. Additionally, PASS exhibits strong robustness against variations in the service area along the waveguide direction. These findings reveal the great potential of PASS in the next-generation wireless networks. Future works may consider more computationally efficient PA placement optimization algorithms.

\appendices
\section{Proof of Theorem 1} \label{proof_of_theorem1}
We proof Theorem \ref{theorem1} by using a contradiction following \cite{Obiedollah2025energy}. Let $\mathcal{F}({\cdot})$ denote the objective function of problem (\textbf{P1-1}). Assume that there is another solution $(\bm{{\rm w}}', \bm{{\rm X}}')$ that contradicts the Pareto-optimal condition, that is, under the solution $(\bm{{\rm w}}', \bm{{\rm X}}')$, one objective (SE or EE) can be improved without degrading the other, which can be expressed as
\begin{equation}\label{eq_theorem1_proof_1}
	\setlength\abovedisplayskip{3pt}
	\setlength\belowdisplayskip{3pt}
	f_{{\rm SE}}(\bm{{\rm w}}', \bm{{\rm X}}') \geq f_{{\rm SE}}(\bm{{\rm w}}^*, \bm{{\rm X}}^*),
\end{equation}
\begin{equation}\label{eq_theorem1_proof_2}
	\setlength\abovedisplayskip{3pt}
	\setlength\belowdisplayskip{3pt}
	f_{{\rm EE}}(\bm{{\rm w}}', \bm{{\rm X}}') \geq f_{{\rm EE}}(\bm{{\rm w}}^*, \bm{{\rm X}}^*).
\end{equation}
Denote $v_{{\rm SE}}^* = [f_{{\rm SE}}(\bm{{\rm w}}^*, \bm{{\rm X}}^*)]^{\beta}$, $v_{{\rm EE}}^* = [f_{{\rm EE}}(\bm{{\rm w}}^*, \bm{{\rm X}}^*)]^{1-\beta}$, $v_{{\rm SE}}' = [f_{{\rm SE}}(\bm{{\rm w}}', \bm{{\rm X}}')]^{\beta}$, and $v_{{\rm EE}}' = [f_{{\rm EE}}(\bm{{\rm w}}', \bm{{\rm X}}')]^{1-\beta}$. Since $(\bm{{\rm w}}^*, \bm{{\rm X}}^*)$ is the optimal solution for SOO problem (\textbf{P1-1}), the following condition holds for any feasible $(\bm{{\rm w}}', \bm{{\rm X}}')$:
\begin{equation}
	\setlength\abovedisplayskip{3pt}
	\setlength\belowdisplayskip{3pt}
	v_{{\rm SE}}^* v_{{\rm EE}}^* \geq v_{{\rm SE}}' v_{{\rm EE}}'.
\end{equation}
The above condition can be equivalently expressed as\footnote{Noted that $\beta\geq 0$, and SE and EE are greater than 0.}
\begin{equation}\label{eq_theorem1_proof_4}
	\frac{{v_{{\rm SE}}^*}}{v_{{\rm SE}}'} \geq \frac{v_{{\rm EE}}'}{v_{{\rm EE}}^*}.
\end{equation}
As a consequence, there are two cases to be considered: 

Case 1: $v_{{\rm EE}}'/v_{{\rm EE}}^*\geq 1$. It means that $f_{{\rm EE}}(\bm{{\rm w}}', \bm{{\rm X}}') \geq f_{{\rm EE}}(\bm{{\rm w}}^*, \bm{{\rm X}}^*)$. According to (\ref{eq_theorem1_proof_4}), we have ${v_{{\rm SE}}^*}/v_{{\rm SE}}' \geq 1$, which implies that $f_{{\rm SE}}(\bm{{\rm w}}^*, \bm{{\rm X}}^*) \geq f_{{\rm SE}}(\bm{{\rm w}}', \bm{{\rm X}}')$. Therefore, the conditions (\ref{eq_theorem1_proof_1}) and (\ref{eq_theorem1_proof_2}) cannot be satisfied simultaneously.

Case 2: $v_{{\rm EE}}'/v_{{\rm EE}}^*\leq 1$. It implies that $f_{{\rm EE}}(\bm{{\rm w}}', \bm{{\rm X}}') \leq f_{{\rm EE}}(\bm{{\rm w}}^*, \bm{{\rm X}}^*)$ which breaks the condition (\ref{eq_theorem1_proof_2}).

Based on the above analysis, it is impossible to have any feasible $(\bm{{\rm w}}', \bm{{\rm X}}')$ such that $\mathcal{F}(\bm{{\rm w}}', \bm{{\rm X}}')\geq \mathcal{F}(\bm{{\rm w}}^*, \bm{{\rm X}}^*)$. This result imposes that the solution $(\bm{{\rm w}}^*, \bm{{\rm X}}^*)$ is a Perto-optimal solution, since there is no other solution that can achieve $\mathcal{F}(\bm{{\rm w}}', \bm{{\rm X}}')\geq \mathcal{F}(\bm{{\rm w}}^*, \bm{{\rm X}}^*)$. The proof is thus completed.

\section{Proof of Lemma 2} \label{proof_of_lemma2}
Take the first-order derivative of $f_{{\rm EE}}(P)$ as
\begin{equation}\label{proof_lemma2_eq1}
	\setlength\abovedisplayskip{3pt}
	\setlength\belowdisplayskip{3pt}
	[f_{{\rm EE}}(P)]' \!=\! \frac{\zeta(P\!+\!P_f)\!-\!(1+\zeta P){\rm ln}(1\!+\!\zeta P)}{(1\!+\!\zeta P)\left(P+P_f+\frac{\chi}{{\rm ln}2}{\rm ln}(1+\zeta P)\right)^2{\rm ln}2}.
\end{equation}

We notice that the denominator of $[f_{{\rm EE}}(P)]'$ is positive. For the numerator, letting $g(P)\!=\!\zeta(P\!+\!P_f)\!-\!(1\!+\!\zeta P){\rm ln}(1\!+\!\zeta P)$, we have $g'(P)=-\zeta{\rm ln}(1+\zeta P)<0$, implying that $g(P)$ is decreasing with $P$. Furthermore, $\lim_{P\to 0^+}g(P)=\zeta P_f>0$, $\lim_{P\to+\infty}g(P)=-\infty$. Therefore, the equation $g(P)=0$ has a unique root on $[0, +\infty)$, denoted as $P^*$, and $f_{{\rm EE}}(P)$ is monotonically increasing on $[0, P^*)$, and monotonically decreasing on $[P^*, +\infty)$. Meanwhile, letting $l(p)$ represent the denominator part of (\ref{proof_lemma2_eq1}), we can obtain the second-order derivative of $f_{{\rm EE}}(P)$ with $P$ as
\begin{equation}\label{proof_lemma2_eq2}
	\setlength\abovedisplayskip{3pt}
	\setlength\belowdisplayskip{3pt}
	[f_{{\rm EE}}(P)]'' = \frac{g'(P)l(P)-g(P)l'(P)}{(l(P))^2}.
\end{equation}

It is noticed that $l(P)>0$ and $l'(P)>0$ for $\forall P\in [0, +\infty)$. When $P\in [0, P^*]$, we have $g(P)>0$ and $g'(P)<0$, leading to the fact that $[f_{{\rm EE}}(P)]''<0$, which means that $f_{{\rm EE}}(P)$ is concave. While for the interval $P\in (P^*, +\infty)$, $[f_{{\rm EE}}(P)]''$ can be either positive or negative, thus $f_{{\rm EE}}(P)$ is only quasi-concave, and neither concave nor convex. In conclusion, $f_{{\rm EE}}(P)$ is strictly increasing and concave at $[0, P^*]$ while strictly decreasing and only quasi-concave at $(P^*, +\infty)$. The proof is thus completed.

\section{Proof of Proposition 1}\label{proof_of_proposition1}
First, when $P_T \leq P^*$, we know that $f_{{\rm SE}}(P)$ is increasing at $[0, P_T]\in[0, P^*]$ given the fact that $f_{{\rm SE}}(P)$ is strictly increasing with $P$. In addition, according to \textbf{Lemma} \textbf{\ref{lemma2}}, we obtain that $f_{{\rm EE}}(P)$ is also increasing at $[0, P_T]$. Thus, for $\forall P\in [0, P_T)$, we have $f_{{\rm SE}}(P_T) > f_{{\rm SE}}(P)$ and $f_{{\rm EE}}(P_T) > f_{{\rm EE}}(P)$, which results in $\mathcal{P}=\{P|P=P_T\}$.

Next, consider the case when $P_T > P^*$. As mentioned above, for $\forall P\in [0, P^*)$, we can always find a point $P^*$ such that $f_{{\rm SE}}(P^*) > f_{{\rm SE}}(P)$ and $f_{{\rm EE}}(P^*) > f_{{\rm EE}}(P)$, which implies that $[0, P^*) \notin \mathcal{P}$. On the other hand, for $\forall P\in [P^*, P_T]$, there does not exist any other point $P'$ such that both $f_{{\rm SE}}(P') > f_{{\rm SE}}(P)$ and $f_{{\rm EE}}(P') > f_{{\rm EE}}(P)$. Therefore, $\mathcal{P}=\{P|P^* \leq P \leq P_T\}$. The proof is thus completed. 

\section{Proof of Lemma 3}\label{proof_of_lemma3}
First, $g_2(P)$ can be rewritten in eq. (\ref{sec2D_eq7}). It is noticed that the second and third terms of $g_2(P)$ is strictly decreasing with $P$. For the first term, let $A=\frac{\chi \zeta}{{\rm ln}2}$, its first derivative can be given in eq. (\ref{sec2D_eq8}). For the molecule of eq. (\ref{sec2D_eq8}), since $\zeta>0$, we therefore let
\begin{equation}\label{proof_lemma3_eq1}
	\setlength\abovedisplayskip{3pt}
	\setlength\belowdisplayskip{3pt}
	h(P)=(1+A){\rm ln}(1+\zeta P)-\zeta P-\frac{\zeta AP}{1+\zeta P}.
\end{equation}
Furthermore, by defining $\rho=\zeta P$, so we have $\rho>0$, and $h(\rho) = (1+A){\rm ln}(1+\rho)-\rho-A\frac{\rho}{1+\rho}$. The first derivative of $h(\rho)$ can be given as
\begin{equation}\label{proof_lemma3_eq2}
	\setlength\abovedisplayskip{3pt}
	\setlength\belowdisplayskip{3pt}
	h'(\rho)=\frac{1+A}{1+\rho}-1-\frac{A}{(1+\rho)^2}=\frac{\rho[(A-1)-\rho]}{(1+\rho)^2}.
\end{equation}
It is noticed that if $A\leq 1+\rho$, which is a mild requirement in real environments, then $h'(\rho)<0$, which means that $h(P)$ is decreasing with $P$. In addition, $\lim_{P\to 0^+}h(P)=0$, so we can infer that when $P\in [0, +\infty)$, $h(P)\leq 0$. Therefore, we have $\left[\frac{\zeta P}{\left(1+\zeta P + A\right){\rm ln}(1+\zeta P)}\right]'<0$. Thus, the first term of $g_2(P)$ is also decreasing with $P$, so we achieve that $g_2(P)$ is strictly decreasing with $P$. The proof is thus completed.

\begin{IEEEbiography}
[{\includegraphics[width=1in,height=1.25in,clip,keepaspectratio]{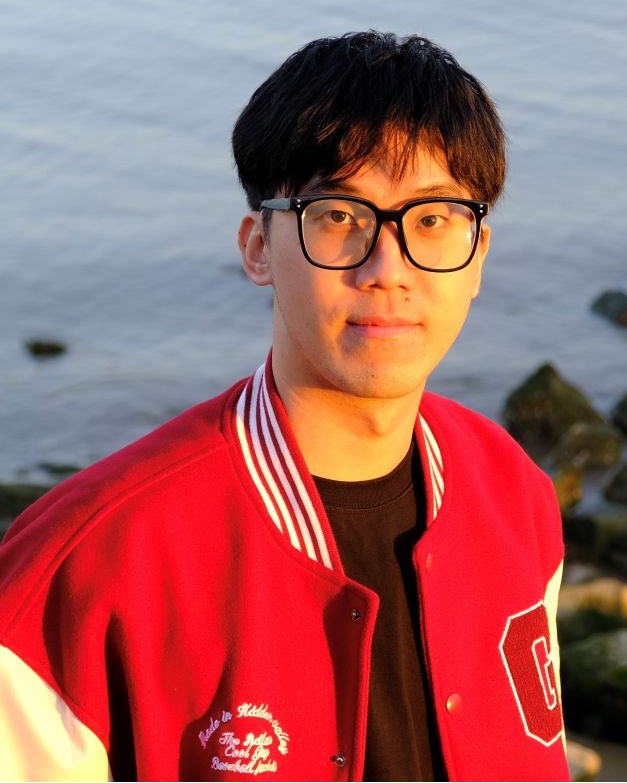}}]{Zihao Zhou}
(Graduate Student Member, IEEE) received the B.Eng. and M. Eng. degrees from the South China University of Technology (SCUT), Guangzhou, China, in 2022 and 2025, respectively. He is currently pursuing the Ph.D. degree with the Department of Electrical and Computer Engineering (ECE), The University of Hong Kong (HKU), Hong Kong SAR, China. His research interests include next-generation wireless systems and AI for communications and networks.
\end{IEEEbiography}

\begin{IEEEbiography}
[{\includegraphics[width=1in,height=1.25in,clip,keepaspectratio]{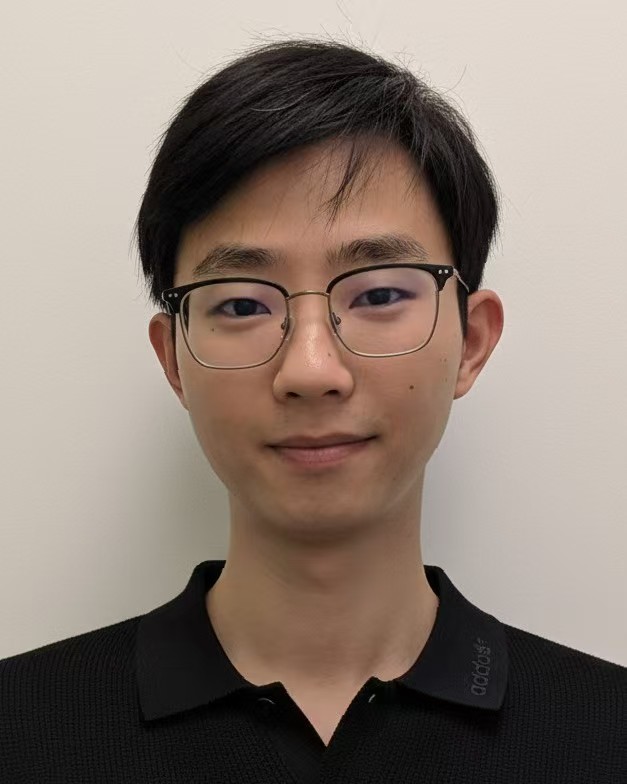}}]{Zhaolin Wang}
(Member, IEEE) received the dual B.Eng. degrees (with honors) from Beijing University of Posts and Telecommunications, China, and Queen Mary University of London (QMUL), U.K., in 2020, the M.Sc. degree (with distinction) from Imperial College London, U.K., in 2021, and the Ph.D. degree from QMUL in 2024.

He is currently a Research Assistant Professor at The University of Hong Kong. Prior to that, he was a Postdoctoral Researcher with QMUL from 2024 to 2025. His research focuses on electromagnetic signal and information theory, integrated sensing and communications, and artificial intelligence for next-generation wireless systems. He has received the Best Student Paper Award at IEEE VTC2022-Fall, the 2023 IEEE Daniel E. Noble Fellowship Award, and the 2025 Andrea Goldsmith Young Scholar Award. He was also recognized as an Exemplary Reviewer for IEEE Wireless Communications Letters in 2023 and IEEE Communications Letters in 2024. He currently serves as an Editor for IEEE Transactions on Communications. More information can be found at \url{https://zhaolin820.github.io}.
\end{IEEEbiography}

\begin{IEEEbiography}
[{\includegraphics[width=1in,height=1.25in,clip,keepaspectratio]{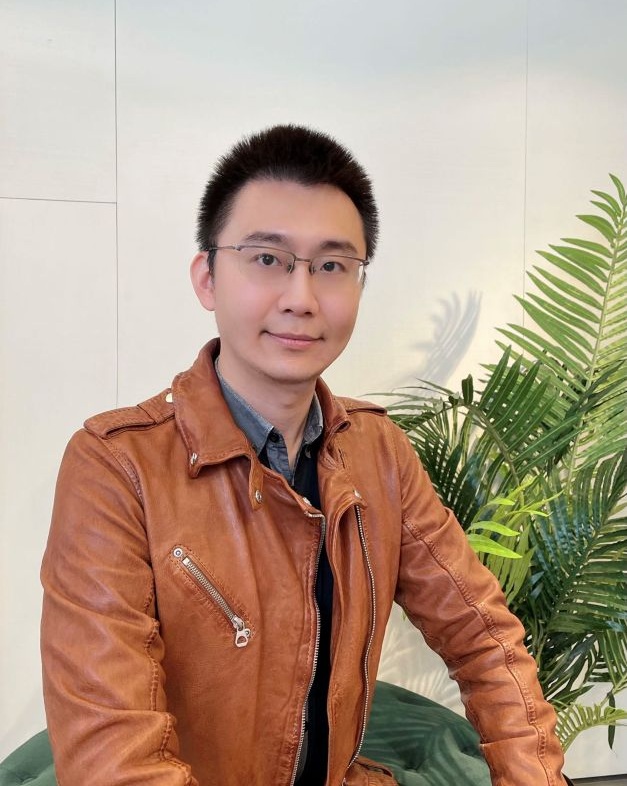}}]{Yuanwei Liu}
(Fellow, IEEE) received the Ph.D. degree from the Queen Mary University of London (QMUL), London, U.K., in 2016. He is currently a tenured Full Professor with the Department of Electrical and Computer Engineering (ECE), The University of Hong Kong (HKU). Prior to that, he was a Senior Lecturer (Associate Professor) (2021--2024) and a Lecturer (Assistant Professor) (2017--2021) at QMUL and a Post-Doctoral Research Fellow (2016--2017) at King's College London (KCL), London. His research interests include generative AI/LLM for communications, low altitude networks large model, mobile edge generation, wireless digital twins, and integrated sensing and communications. He is a fellow of AAIA, a Web of Science Highly Cited Researcher, a Young Member of Hong Kong Academy of Engineering, an IEEE Communication Society Distinguished Lecturer, an IEEE Vehicular Technology Society Distinguished Lecturer, and the Rapporteur of ETSI Industry Specification Group on Reconfigurable Intelligent Surfaces on Work Item of Multi-Functional Reconfigurable Intelligent Surfaces (RIS): Modeling, Optimization, and Operation. He serves the Chair for IEEE Signal Processing and Computing for Communications (SPCC) and the Academic Chair for the Next Generation Multiple Access Emerging Technology Initiative. He was listed as one of 35 Innovators Under 35 China in 2022 by MIT Technology Review. He received IEEE ComSoc Outstanding Young Researcher Award for EMEA in 2020. He received the 2020 IEEE SPCC Technical Committee Early Achievement Award and IEEE Communication Theory Technical Committee (CTTC) 2021 Early Achievement Award. He received IEEE ComSoc Outstanding Nominee for Best Young Professionals Award in 2021. He was a co-recipient of the 2024 IEEE Communications Society Heinrich Hertz Award, the Best Student Paper Award in IEEE VTC2022-Fall, the Best Paper Award in ISWCS 2022, the 2022 IEEE SPCCTC Best Paper Award, the 2023 IEEE ICCT Best Paper Award, and the 2023 IEEE ISAP Best Emerging Technologies Paper Award. He serves as the Co-Editor-in-Chief for IEEE ComSoc TC Newsletter, an Area Editor for \textsc{IEEE Transactions on Communications} and \textsc{IEEE Communications Letters}, an Editor for \textsc{IEEE Communications Surveys \& Tutorials}, \textsc{IEEE Transactions on Wireless Communications}, \textsc{IEEE Transactions on Vehicular Technology}, \textsc{IEEE Transactions on Network Science and Engineering}, and \textsc{IEEE Transactions on Cognitive Communications and Networking}. He serves as the (leading) Guest Editor for \textsc{Proceedings of the IEEE} on Next Generation Multiple Access, \textsc{IEEE Journal on Selected Areas in Communications} on Next Generation Multiple Access, \textsc{IEEE Journal of Selected Topics in Signal Processing} on Intelligent Signal Processing and Learning for Next Generation Multiple Access, and \textsc{IEEE Network} on Next Generation Multiple Access for 6G. He serves as the Publicity Co-Chair for IEEE VTC 2019-Fall, the Panel Co-Chair for IEEE WCNC 2024, and the Symposium Co-Chair for several flagship conferences, such as IEEE GLOBECOM, ICC, and VTC.
\end{IEEEbiography}

\vspace{11pt}

\newpage

\vspace{11pt}

\vfill

\end{document}